%% file: HeteroticOrbifoldModelsArXiV.tex
\documentclass[fleqn,11pt]{article} 
\usepackage{newtxtext,newtxmath} 
\usepackage{mathrsfs}
\usepackage{graphicx}        
\usepackage[bottom]{footmisc}
\usepackage[colorlinks=false]{hyperref}        
\usepackage{soul}            
\usepackage{microtype}       
\hypersetup{colorlinks=true,urlcolor=blue,citecolor=orange}
\usepackage[square,numbers]{natbib}
\usepackage[textsize=tiny,backgroundcolor=yellow]{todonotes}
\usepackage{dsfont}%
\usepackage{booktabs}%
\usepackage{cancel}%
\usepackage{subcaption}%
\usepackage[inline]{enumitem}%
\usepackage[centercolon]{mathtools}
\usepackage{cleveref}
\usepackage[nolist]{acronym}
\usepackage{nicefrac}
\usepackage{xspace}
\usepackage{mleftright}
\mleftright
\usepackage{array}
\usepackage{geometry}
\makeatletter
\g@addto@macro\bfseries{\boldmath}
\makeatother
\newlist{goals}{enumerate*}{1}
\setlist[goals]{label=(\arabic*),ref=(\arabic*)}
\crefname{goalsi}{goal}{goals}
\Crefname{goalsi}{Goal}{Goals}
\newlist{challenges}{enumerate}{1}
\setlist[challenges]{label=C\arabic*.,ref=C\arabic*,labelindent=\parindent,leftmargin=*}
\crefname{challengesi}{}{}
\Crefname{challengesi}{}{}
\newcommand{\ii}{\mathop{}\!\mathrm{i}\!\mathop{}}

\newcommand{\CP}{\ensuremath{\mathcal{CP}}\xspace}

\newcommand{\WriteUp}{review\xspace}

\begin{document}
\title{\Huge\bfseries Heterotic Orbifold Models}
\author{Sa\'ul Ramos-S\'anchez\textsuperscript{1}\thanks{ramos@fisica.unam.mx}~~ and
	Michael Ratz\textsuperscript{2}\thanks{mratz@uci.edu}}%
\date{\begin{tabular}{>{\centering}p{0.95\textwidth}}
  \begin{tabular}{>{\centering\small}p{0.9\textwidth}}
 {}\textsuperscript{1}\,Universidad Nacional Aut\'onoma de M\'exico,\\ POB 20-364, Cd.Mx. 01000, M\'exico\\[2em]  
 {}\textsuperscript{2}\,Department of Physics and Astronomy,\\
 University of California,\\ Irvine, CA 92697-4575, USA
\end{tabular}\\[2em] 
\em Invited chapter for the Handbook of Quantum Gravity (edited by Cosimo Bambi, Leonardo Modesto, and Ilya Shapiro, Springer 2023)
\end{tabular}}
\maketitle
\abstract{%
We review efforts in string model building, focusing on the heterotic orbifold compactifications.
We survey how one can, starting from an explicit string theory, obtain models which resemble Nature. 
These models exhibit the standard model gauge group, three generations of standard model matter and an appropriate Higgs sector.
Unlike many unified models, these models do not suffer from problems such as doublet-triplet splitting, too rapid proton decay and the $\mu$ problem.
Realistic patterns of fermion masses emerge, which are partly explained by flavor symmetries, including their modular variants. 
We comment on challenges and open questions.
}

\section*{Keywords}
String compactifications, Heterotic string, model building,
phenomenology, symmetries of string models.
\clearpage
\markboth{Heterotic Orbifold Models}{Heterotic Orbifold Models}

\acresetall

\section{Introduction}

If string theory is to describe the real world, it has to reproduce our current established understanding of physics. 
In particular, its low-energy description has to give rise to the \ac{SM}. 
Generally, string model building concerns the question of how the \ac{SM} fits into string theory. 
In practice, one compactifies a consistent string theory to a four-dimensional model which can be studied and confronted with observation. 
One particularly important aspect of top-down model building is that the globally consistent models are complete, i.e.\ they do not only describe the \ac{SM} but also include, say, the degree(s) of freedom driving cosmic inflation and dark matter. 
That is, unlike in the bottom-up approach, one cannot add extra sectors to the model at will.

Historically, the first attempts to construct realistic string models were based on the heterotic string. 
It was noticed that the structure of \ac{SM} is remarkably consistent with unification along the exceptional chain $\text{SU}(5)\subset\text{SO}(10)\subset\text{E}_6\subset\text{E}_7\subset\text{E}_8$~\cite{Olive:1981tb}. 
In this \WriteUp we provide a brief overview of orbifold compactifications of the heterotic string which come close to the \ac{SM}.

\begin{figure}[htb]
 \centering
 \includegraphics[width=\linewidth]{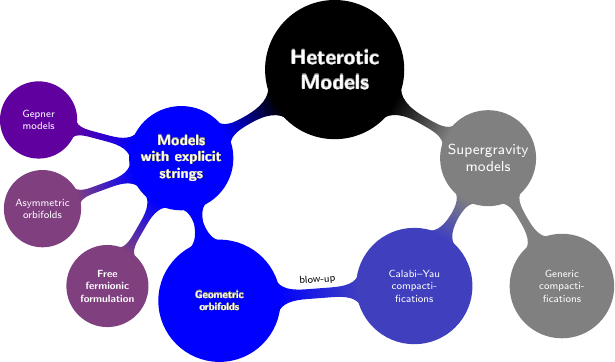}
 \caption{An incomplete survey of heterotic models. The focus of this
 \WriteUp is on the constructions that are typeset bold. 
 \label{fig:HeteroticModels_Mindmap}}
\end{figure}

Heterotic models come broadly in two classes, they can either be based on smooth compactifications~\cite{Candelas:1985en} or on obifolds~\cite{Dixon:1985jw,Dixon:1986jc}, cf.\ \Cref{fig:HeteroticModels_Mindmap}.  These classes are related as some smooth compactifications can emerge from orbifolds via blow-up (cf.\ e.g.\ \cite{Blaszczyk:2010db}).
Orbifolds have the advantage that their construction involves explicit strings, which is why they will be our focus. 
Orbifolds can be constructed in the so-called free fermionic approach, yet our focus will be the classical approach, in which the so-called symmetric orbifolds have a geometric interpretation.

The purpose of this \WriteUp is to summarize the current status of heterotic model building. 
There are already excellent reviews of this subject such as \cite{Ibanez:2012zz}, however, our focus will be on more recent developments and a clear account of the open questions. 
To this end, we will review the target of string model building, the \ac{SM} and some of its extensions in \Cref{sec:PhenomenologicalConstraints} before turning to the heterotic string and its compactifications in \Cref{sec:Compactifications}. 
In \Cref{sec:Spectrum,sec:Symmetries} we collect some facts about the spectra and symmetries of the constructions, which will be the basis for the discussion of challenges in \Cref{sec:Challenges}. 
In \Cref{sec:Examples} we provide some explicit examples. 
After briefly commenting on smooth heterotic compactifications in \Cref{sec:Smooth_Compactifications}, we provide an outlook in \Cref{sec:HeteroticModels_Outlook}.

\section{What do we (believe to) know?}
\label{sec:PhenomenologicalConstraints}

Before delving into what string theory gives us, let us briefly survey what we expect to get out of string model building.

\subsection{A very short recap of the \ac{SM}}

First and foremost, we wish to obtain a \ac{QFT} that is consistent with the \ac{SM} (see e.g.\ \cite{Langacker:2017uah} for a detailed description). 
The latter is based on the continuous gauge symmetry\footnote{Strictly speaking we do not really know the gauge group of the \ac{SM} but only its Lie algebra, a subtlety which we will, like most of the literature, not discuss in detail.}
\begin{equation}\label{eq:C_SM}
 G_\text{SM}=\text{SU}(3)_\text{C}\times\text{SU}(2)_\text{L}\times\text{U}(1)_\text{Y}\;.
\end{equation}
The matter content consists of three generations of quarks and leptons, left-chiral Weyl fermions which transform as 
\begin{subequations}\label{eq:quarks_and_leptons}
\begin{align}
 \text{quarks }&:~q_f=(\boldsymbol{3},\boldsymbol{2})_{\nicefrac{1}{6}}
 \;,~\bar u_f=(\overline{\boldsymbol{3}},\boldsymbol{1})_{-\nicefrac{2}{3}}
 \;,~\bar d_f=(\overline{\boldsymbol{3}},\boldsymbol{1})_{\nicefrac{1}{3}}
 \;,\\
 \text{leptons }&:~\ell_f=(\boldsymbol{1},\boldsymbol{2})_{-\nicefrac{1}{2}}
 \;,~\bar e_f=(\boldsymbol{1},\boldsymbol{1})_{1}
\end{align}
\end{subequations}
under $G_\text{SM}$. 
Here, $f\in\{1,2,3\}$ labels the generations. 
In addition, there is the so-called Higgs field, a complex scalar carrying the quantum numbers $h=(\boldsymbol{1},\boldsymbol{2})_{\nicefrac{1}{2}}$. 
The Higgs acquires a \ac{VEV}, $\langle h\rangle\sim100\,\text{GeV}$, which breaks $G_\text{SM}$ down to $\text{SU}(3)_\text{C}\times\text{U}(1)_\text{em}$, under which \eqref{eq:quarks_and_leptons} are vector-like and acquire masses which are given by the product of so-called Yukawa couplings $Y_{u,d,e}$ and $\langle h\rangle$. 
In the \ac{SM}, the Yukawa couplings are input parameters which are adjusted to fit data. 
A curious fact about the \ac{SM} is that the combination of charge conjugation and parity, $\mathcal{CP}$, is broken in the flavor sector, i.e.\ by the Yukawa couplings, but seemingly not in the strong interactions. 
This mismatch gets referred to as the strong $\mathcal{CP}$ problem. 
The neutrinos, which are part of the $\ell_f$, are also massive, yet it is currently not known which operator describes their mass. 
The most plausible options are the Weinberg operator, $\kappa_{gf}(\ell^g h)(\ell^f h)$, or Dirac neutrino masses, in which case one has to amend \eqref{eq:quarks_and_leptons} by right-handed neutrinos $\nu^f$. 
The neutrino masses are much smaller than the masses of the charged fermions.

It is instructive to survey the continuous parameters of the \ac{SM}. They comprise
\begin{enumerate*}[label=(\roman*)]
 \item $3$ gauge couplings;
 \item $\theta_\mathrm{QCD}$;
 \item $2$ Higgs parameters;
 \item $12$ masses;
 \item $8$ or $10$ mixing parameters, depending on whether neutrinos are Dirac  or Majorana particles.
\end{enumerate*}
In a stringy completion, these parameters should be predicted rather than adjusted, and, as we shall discuss in \Cref{sec:Challenges}, the requirement to reproduce these observables remains one of the greatest challenges in string model building. 
Note also that currently the only other parameters that we need to describe observation are the Planck mass  $M_\text{P}$ (or, equivalently, $G_\text{Newton}$), the vacuum energy  $\rho_\text{vacuum}$ and the density to dark matter,  $\rho_\text{DM}$. 
That is, currently the bulk of the (ununderstood) parameters of Nature resides in the flavor sector of the \ac{SM}.

An important fact about the \ac{SM} is that it does not only provide us with couplings and interactions that have been confirmed in experiments, but it also comes with tight constraints on additional particles and interactions. 
In particular, it is extremely hard to make extra states which are chiral w.r.t.\ $G_\text{SM}$ consistent with observation. 
Also, while the Weinberg operator is a nonrenormalizable operator that is ``good'' in the sense that it can describe neutrino masses, other higher-dimensional operators are highly constrained. 
For instance, the suppression scale of the dimension-6 operators leading to proton decay has to exceed $10^{15}\,\text{GeV}$.

\subsection{Early universe}
\label{subsec:EarlyUniverse}

The early universe provides us with important insights into high energy physics (see e.g.\ \cite{Baumann:2022mni}). 
For instance, \ac{BBN} works very well within the \ac{SM}, and extra particles may be inconsistent with the primordial formation of the elements if they decay late or increase the Hubble expansion rate too much. 
In addition, fractionally charged particles are often stable since they cannot decay into \ac{SM} states, and there are stringent constraints on their relic abundance (cf.\ e.g.\ \cite{Langacker:2011db,Halverson:2018vbo}).

However, the early universe also requires physics beyond the \ac{SM}. 
Most notably we need a field or sector that drive inflation, or another ingredient which provides us with solutions to the so-called horizon and flatness problems. 
In addition, there is very compelling evidence for dark matter which cannot be made of \ac{SM} particles. 
Furthermore, the baryon asymmetry of the universe requires physics beyond the \ac{SM}, too. 

\subsection{\Ac{BSM} scenarios}
\label{subsec:BSM}

Having seen that physics beyond the \ac{SM} is required to accommodate astrophysics and cosmology, let us spend some words on \ac{BSM} scenarios.

The supersymmetric variants of the \ac{SM} (see e.g.\ \cite{Martin:1997ns} for an introduction), most notably the \ac{MSSM}, have received substantial attention in the past decades. 
This is because, assuming low-energy \ac{SUSY}, the electroweak scale gets stabilized against quantum corrections. 
Of course, given the absence of clear signals for \ac{SUSY} at the \ac{LHC}, this scheme has lost some of its popularity in the recent years, yet the \ac{MSSM} is arguably still one of the best motivated and well-defined \ac{BSM} scenarios. 
The \ac{MSSM} has a number of shortcomings which one may hope to solve in \ac{UV} completions, and the purpose of this \WriteUp is to discuss these solutions in \Cref{sec:Examples}. 
In order to understand some of these shortcomings, let us look at the $G_\text{SM}$ invariant superpotential terms up to  order 4,
\begingroup\let\SuperField\empty
\newcommand*{\SFConjugate}[1]{\overline{#1}}%
\begin{align}
\MoveEqLeft\mathscr{W}_\text{gauge~invariant} =
\mu\, \SuperField{h}_d \SuperField{h}_u +
\kappa_i\, \SuperField{\ell}_i
\SuperField{h}_u
\nonumber \\
& \quad{}
+ Y_e^{gf}\, \SuperField{\ell}_g \SuperField{h}_d \SFConjugate{\SuperField{e}}_f
+ Y_d^{gf}\, \SuperField{q}_g \SuperField{h}_d \SFConjugate{\SuperField{d}}_f
+ Y_u^{gf}\, \SuperField{q}_g \SuperField{h}_u \SFConjugate{\SuperField{u}}_f
\nonumber\\
& \quad {} +
\lambda_{gfk}\, \SuperField{\ell}_g \SuperField{\ell}_f \SFConjugate{\SuperField{e}}_k
+
\lambda^\prime _{gfk}\, \SuperField{\ell}_g \SuperField{q}_f \SFConjugate{\SuperField{d}}_k
+
\lambda^{\prime\prime}_{gfk}\, \SFConjugate{\SuperField{u}}_g\SFConjugate{\SuperField{d}}_f\SFConjugate{\SuperField{d}}_k
\nonumber\\
& \quad {}
+\kappa_{gf}\, \SuperField{h}_u \SuperField{\ell}_g\,\SuperField{h}_u
\SuperField{\ell}_f
+
\kappa^{(1)}_{gfk\ell}\, \SuperField{q}_g \SuperField{q}_f \SuperField{q}_k \SuperField{\ell}_\ell
+
\kappa^{(2)}_{gfk\ell}\,
\SFConjugate{\SuperField{u}}_g\SFConjugate{\SuperField{u}}_f\SFConjugate{\SuperField{d}}_k \SFConjugate{\SuperField{e}}_\ell
\;,\label{eq:W_MSSM}
\end{align}
\endgroup
where we, in a slight abuse of notation, denoted the superfields by the same symbols as the \ac{SM} fields in \Cref{eq:quarks_and_leptons}. 
Note also that the \ac{MSSM} has two Higgs doublets, $h_u$ and $h_d$. 
The couplings $Y_u$, $Y_d$ and $Y_e$ are the Yukawa couplings which yield the masses of quarks and charged leptons. 
The $R$-parity violating terms $\kappa_i$, $\lambda_{gfk}$, $\lambda_{gfk}^{\prime}$ and $\lambda_{gfk}^{\prime\prime}$ have to be highly suppressed, and get often forbidden by $R$ (or matter) parity, the origin of which is to be clarified in a \ac{UV} completion of the model. 
The $\mu$ term in the first line of \Cref{eq:W_MSSM} can a priori have any size, but in order to have proper electroweak symmetry breaking and sufficiently heavy Higgsinos it should be of the order TeV or so, which is a common choice for the soft \ac{SUSY} breaking masses. 
Explanations of this fact comprise the Kim--Nilles \cite{Kim:1983dt} and Giudice--Masiero \cite{Giudice:1988yz} mechanisms, and we will see later in \Cref{sec:Examples} both are realized in explicit stringy completions of the \ac{MSSM}. 
Further, while the $\kappa$ term in the last line of \Cref{eq:W_MSSM} can describe neutrino masses, the $\kappa^{(i)}$ terms have to be very small, e.g.\ the coefficients $\kappa^{(1)}_{1121}$ and $\kappa^{(1)}_{1122}$ have to be suppressed by more than $10^8\cdot M_\text{P}$, where $\sqrt{8\pi}\,M_\text{P}=M_\text{Planck}\simeq1.2\cdot10^{19}\,\text{GeV}$. 
A proper understanding of this suppression arguably requires a solution within a consistent theory of quantum gravity, such as the explicit string models we consider here.

Another appealing feature of the \ac{MSSM} with low-energy \ac{SUSY} is that gauge couplings unify remarkably well at a scale $M_\text{GUT}\simeq\text{few}\times10^{16}\,\text{GeV}$ \cite{Dimopoulos:1981yj}. 
This has led to the scheme of \ac{SUSY} \acp{GUT} (see e.g.\ \cite{Raby:2017ucc} for an extended discussion), in which a unified symmetry like $\text{SU}(5)$ or $\text{SO}(10)$ gets broken at $M_\text{GUT}$ down to $G_\text{SM}$. 
An arguably even stronger motivation for \acp{GUT} is the structure of matter since one generation of the \ac{SM} (cf.\ \Cref{eq:quarks_and_leptons}) including the right-handed neutrino its into a $\boldsymbol{16}$-plet of $\text{SO}(10)$,
\begin{equation}\label{eq:MSSM_GUT}
 \boldsymbol{16}=
 \underbrace{(\boldsymbol{3},\boldsymbol{2})_{\nicefrac{1}{6}}
 +(\boldsymbol{\overline{3}},\boldsymbol{1})_{-\nicefrac{2}{3}}
 +(\boldsymbol{1},\boldsymbol{1})_{1}}_{{}={\,}\boldsymbol{10}}
 +\underbrace{(\boldsymbol{\overline{3}},\boldsymbol{1})_{\nicefrac{1}{3}}
 +(\boldsymbol{1},\boldsymbol{2})_{-\nicefrac{1}{2}}}_{{}={\,}\boldsymbol{\overline{5}}}
 +(\boldsymbol{1},\boldsymbol{1})_{0}\;,
\end{equation}
where we also indicated the $\text{SU}(5)$ representations in underbraces. 
While the \ac{GUT} symmetries work very well for the matter, they fail for the \ac{SM} Higgs. 
The smallest $\text{SU}(5)$ representations that contain the \ac{MSSM} Higgs doublets are $\boldsymbol{5}+\overline{\boldsymbol{5}}$, which combine to a $\boldsymbol{10}$-plet of $\text{SO}(10)$. 
The additional $\text{SU}(3)_\text{C}$ $\boldsymbol{3}$-plets contained in these representations pose major threats to the model as they typically mediate unacceptably large proton decay unless their mass exceeds the Planck scale. 
This is one facet of the doublet-triplet splitting problems which haunt \acp{GUT} in 4D. 
On the other hand, as has been pointed out early on in the context of string model building, in higher-dimensional, in particular stringy, models the same mechanism that breaks the \ac{GUT} symmetry can also split the doublets from the triplets \cite{Candelas:1985en,Breit:1985ud}.

\subsection{What do we hope to learn from string model building?}

Of course, it will be reassuring to find a compactification which reproduces the \ac{SM} in great detail, regardless of whether or not the underlying construction is unique. 
What is more, given such a model, we will be in a unique position to answer some of the most popular questions of our time:
\begin{enumerate}
 \item What is the origin of flavor and $\mathcal{CP}$ violation?
 \item What is the nature of dark matter and what are the properties of the dark (aka hidden) sector?
 \item What drives cosmic inflation?
\end{enumerate}
While these questions might be answered separately, the power of addressing them in explicit string models is that the answers are much more specific and related in intriguing ways.

\section{Compactifying the heterotic string}
\label{sec:Compactifications}

\subsection{Heterotic string}
\label{sec:HetString}

In this section, we collect some basic facts on the heterotic string.
For further details and a broader overview see~\cite{HandbookChapter1arxiv}.
The term `heterotic' derives from the Greek word `hetero', which translates as `other', and in biology is related to `vigorous hybrid', which arguably reflects the nature of the heterotic string. 
The heterotic string theory \cite{Gross:1984dd} is the result of combining a 10D superstring and a 26D bosonic string. 
The former can equip the theory with $\mathcal{N}=1$ supersymmetry in ten dimensions whereas the bosonic string provides us with a non-Abelian gauge group of rank 16,\footnote{The nonsupersymmetric heterotic string can be obtained from this version, as we briefly describe in \Cref{sec:nonSUSY}.}
\begin{equation}\label{eq:G_het}
 G_\text{het}=\text{E}_8\times\text{E}_8\quad\text{or}\quad\text{SO}(32)\;. 
\end{equation}
Note that the most general compactification can have continuous enhancements of these gauge symmetries, yet we will mainly focus on \eqref{eq:G_het}. 
The heterotic theories contain only oriented closed strings propagating in ten dimensions.

In lightcone gauge, there are 8 right-moving bosonic string coordinates $X_\text{R}^i(t-\sigma)$ and 8 right-moving fermions $\psi_\text{R}^i(t-\sigma)$, where $2\le i\le9$. $t$ and $\sigma$ denote the worldsheet coordinates. 
There are in total 24 left-moving coordinates $X_\text{L}^M(t+\sigma)$. In symmetric compactifications they get decomposed into $X_\text{L}^i(t+\sigma)$ with $2\le i\le9$ as in the right-handed  sector, and $X_\text{L}^I(t+\sigma)$ with $1\le I\le16$. 
This decomposition gives rise 8 combinations of ordinary physical coordinates, $X^i(\tau,\sigma)=X_\text{L}^i(t+\sigma)+X_\text{R}^i(t-\sigma)$. 
The additional left-moving coordinates $X_\text{L}^I(t+\sigma)$ are responsible for the gauge symmetries $G_\text{het}$, cf.\ \Cref{eq:G_het}. 
 
For the sake of keeping this \WriteUp short, we specialize on symmetric compactifications, cf.\ \Cref{fig:HeteroticModels_Mindmap}. 
Interestingly, the so-called \ac{FFF} and $\mathds{Z}_2\times\mathds{Z}_2$ geometric orbifolds are related by a dictionary \cite{Donagi:2008xy,Athanasopoulos:2016aws}. 
There are possibilities to go more general, and consider e.g.\ asymmetric orbifolds \cite{Narain:1986qm,Ibanez:1987pj} or Gepner models \cite{Gepner:1987sm,Gepner:1987vz}, which is beyond the scope of this \WriteUp.

\subsection{Heterotic strings on orbifolds}
\label{sec:Strings_on_orbifolds_B}

We will start our discussion with heterotic orbifolds \cite{Dixon:1985jw,Dixon:1986jc}, which allow one to explicitly ``see'' the strings. 
For simplicity, we focus on symmetric toroidal orbifolds, which emerge by dividing tori by some of their symmetries. 
The tori are given by $\mathds{R}^n/\Lambda$ or $\mathds{C}^{n/2}/\Lambda$, where $\Lambda$ denotes a lattice, or, more precisely, the group of lattice translations. We will be interested in $n=6$. 
Therefore, $\mathds{O}=\mathds{R}^n/\mathds{S}$ with $\mathds{S}$ denoting the so-called space group, which is comprised of lattice translations and additional operations such as rotations and so-called roto-translations, and forms a discrete subgroup of the $n$-dimensional Euclidean group. 
Crucially, these operations are also embedded in the gauge sector, which breaks $G_\text{het}$ down to a subgroup. 
Moreover, they also break \ac{SUSY}, which facilitates the construction of chiral 4D models with $\mathcal{N}=1$ or no \ac{SUSY}. 

In more detail, in the geometric formulation elements of space group $\mathds{S}$ are conveniently denoted by $g=(\vartheta^r,m_\alpha e_\alpha)$, where $r,m_\alpha\in\mathds{N}_0$. 
The $e_\alpha$ ($1\le\alpha\le6$) are the basis vectors of the underlying torus. 
The set of $\vartheta\in\text{O}(n)$ form a finite group, called the point group $\mathds{P}$, which determines the holonomy group, and thus the amount of \ac{SUSY} that survives the compactification. 
In fact, in order to classify the physically inequivalent orbifolds, one only needs to find the different affine classes~\cite{Fischer:2012qj}, but we refrain from spelling this discussion out in detail. 
If $\mathds{P}$ is Abelian, it is either $\mathds{Z}_N$ or $\mathds{Z}_N\times\mathds{Z}_M$. If in addition $\mathcal{N}\ge1$ \ac{SUSY} is preserved, then a given $\mathds{Z}_N$ transformation can be encoded in a so-called 3-component twist vector $v$, which describes the rotations of three complex coordinates. 
In general, $g$ acts on string coordinates $X$ as
\begin{equation}
 X\xmapsto{~(\vartheta^r,m_\alpha e_\alpha)~}
 \vartheta^r\,X+m_\alpha e_\alpha\;.
 \label{eq:SpaceGroupActionOnCoordinates}
\end{equation}
The space group is to be embedded in the gauge degrees of freedom. 
Loosely speaking, the point group elements get mapped to so-called shift vectors $V$. 
This embedding has to preserve the order, i.e.\ if $\vartheta\in P$ satisfies $\vartheta^N=\mathds{1}$ then  $N\,V\in\Lambda_{\mathfrak{g}_\text{het}}$, where $\Lambda_{\mathfrak{g}_\text{het}}$ denotes the root lattice of $G_\text{het}$, cf.\ \Cref{eq:G_het}. 
In $\mathds{Z}_N$ orbifolds, if, say, the shifts of two models differ by lattice vectors $\lambda\in\Lambda_{\mathfrak{g}_\text{het}}$, the resulting models are identical (cf.\ \cite{Giedt:2001zw}). 
This is no longer true in $\mathds{Z}_N\times\mathds{Z}_M$ orbifolds, where gauge embeddings differing by lattice vectors may be inequivalent~\cite{Ploger:2007iq}, and be related via what is known as discrete torsion~\cite{Vafa:1986wx}. 
Since $\Lambda_{\mathfrak{g}_\text{het}}$ is even and self-dual, one can find an Euclidean basis in which the lattice vectors are given by
\begin{equation}\label{eq:p_gauge}
 p=(n_1,\dots,n_d)\quad\text{or}\quad
 p=(n_1+\nicefrac{1}{2},\dots,n_d+\nicefrac{1}{2})\;,
\end{equation}
where $n_i\in\mathds{Z}$, $\sum_i^dn_i\in2\mathds{Z}$ and $d\in\{8,16\}$ denotes the dimensions of the Lie algebras $\text{E}_8$ or $\mathfrak{so}(32)$, respectively. 
The gauge embedding of each translation $e_\alpha$ is a so-called discrete Wilson line $W_\alpha$.
The Wilson lines are constrained by geometry. 
In more detail, since lattice vectors get mapped onto each other by the rotations, the analogous relations have to hold for the Wilson lines,
\begin{equation}
    \label{eq:WLconstraints}
      g\,e_\alpha = \sum_{\beta=1}^6 a_\alpha{}^\beta e_\beta \qquad\Longrightarrow\qquad
      W_\alpha = \sum_{\beta=1}^6 a_\alpha{}^\beta W_\beta + \lambda \quad\text{with}\quad 
      \lambda\in\Lambda_{\mathfrak{g}_\text{het}}\;.
\end{equation}
For instance, in a $\mathds{Z}_3$ orbifold plane one has $\vartheta e_1=e_2$ and $\vartheta e_2=-e_1-e_2$. 
Therefore, $W_1\equiv W_2$ and $W_2\equiv -W_1-W_2$, where ``$\equiv$'' means ``equal up to $\lambda\in\Lambda_{\mathfrak{g}_\text{het}}$''. Thus $3W_1\in \Lambda_{\mathfrak{g}_\text{het}}$. 
This generalizes to other geometries, i.e.\ for a given Wilson line $W_\alpha$ there is an integer $M_\alpha$ such that $M_\alpha\,W_\alpha\in\Lambda_{\mathfrak{g}_\text{het}}$, with no summation over $\alpha$. 
As a consequence, the coefficients $a_\alpha{}^\beta$ in \eqref{eq:WLconstraints} are integer. 

In addition, the orbifold parameters and their gauge embeddings must satisfy a series of constraints in order to ensure world-sheet modular invariance, which guarantees the \ac{UV} consistency of the model~\cite{Dixon:1986jc,Vafa:1986wx}. 
For $\mathds{Z}_N$ orbifolds, these conditions take the form~\cite{Ploger:2007iq}
\begin{align}
  N (V^2 - v^2) &= 0\mod2\;, &
  M_\alpha V\cdot W_\alpha &= 0\mod2\;, \nonumber\\
  M_\alpha W_\alpha^2 &= 0\mod2\;, &
  \operatorname{gcd}(M_\alpha,M_\beta) W_\alpha\cdot W_\beta &= 0\mod2\;, 
\end{align}
where no summation over $\alpha$ nor $\beta$ is implied.

\subsection{Classification of toroidal orbifold geometries}
\label{subsec:Classification_of_toroidal_orbifold_geometries}

While early classifications of viable toroidal orbifolds focused on special kinds of lattices~\cite{Bailin:1999nk}, more recently a richer set of possibilities has been uncovered in the $\mathds{Z}_2\times\mathds{Z}_2$ orbifold~\cite{Donagi:2008xy} and generalized to other point groups~\cite{Fischer:2012qj,Fischer:2013qza}. 
Loosely speaking, the new ingredient of the additional possibilities are space groups which contain roto-translations $(\vartheta^r,m_\alpha e_\alpha)\in\mathds{S}$ but $(\vartheta^r,0)\not\in\mathds{S}$ (cf.\ \cite{Hebecker:2004ce}). 
As a consequence, the fundamental group of the orbifolds (and not just the underlying tori) can be nontrivial. 
Among other things, this allows for non-local, or Wilson line, breaking of the gauge symmetry, which is also being utilized in the context of smooth compactifications \cite{Bouchard:2005ag}. 
Another innovation is the consistent construction of non-Abelian orbifolds~\cite{Konopka:2012gy,Fischer:2013qza}. 
In particular, there are 138 Abelian and 331 non-Abelian space groups~\cite{Fischer:2012qj} of toroidal symmetric orbifolds preserving $\mathcal N=1$ \ac{SUSY} in 4D. 
These geometries have been shown to host many models with gauge symmetry and chiral spectrum of the \ac{MSSM}~\cite{Nilles:2014owa,Olguin-Trejo:2018wpw,Parr:2019bta}, yet their detailed phenomenological properties have not been worked out so far.

\subsection{Anisotropic compactifications}

Because of the gauge symmetries of the heterotic string, heterotic models comply well with the idea of grand unification. 
Breaking the \ac{GUT} symmetry via compactification allows one to elegantly avoid the major shortcomings of 4D \acp{GUT}, most notably the doublet-triplet splitting challenge and its associated proton decay problems. 
However, there is a tension between the scale of gauge coupling unification in the \ac{MSSM}, $M_\text{GUT}\simeq\text{few}\cdot10^{16}\text{GeV}$, and typical compactification radii. 
This is because string theory also describes gravity, and the effective 4D Planck mass is sensitive to the volume of compact space. 
In some more detail, Newton's constant $G_\text{N}$ is related to the fine structure `constant' at the \ac{GUT}/compactification scale, $\alpha_\text{\ac{GUT}}$, and the string tension, $\alpha'$, via~\cite{Witten:1996mz} 
\begin{equation}
 G_\text{N}=\frac{\alpha_{\text{\ac{GUT}}}\,\alpha'}{4}
 \quad\text{impying that }G_\text{N}\gtrsim\frac{\alpha_{\text{\ac{GUT}}}^{4/3}}{M_{\text{\ac{GUT}}}^2}    
\end{equation}
for a weakly coupled theory. 
This value of $G_\text{N}$ is too large for typical values of $M_{\text{\ac{GUT}}}$ and $\alpha_\text{\ac{GUT}}$ (cf.\ our ealier discussion around \Cref{eq:MSSM_GUT}). 
There are various proposals to fix this issue (see e.g.\ \cite{Dienes:1996du}). 
The perhaps most ingenious way to address this problem is M-theory \cite{Witten:1996mz}. 
However, the problem can also be ameliorated in anisotropic compactifications \cite[Footnote~3]{Witten:1996mz}. 
A detailed analysis \cite{Hebecker:2004ce} suggests that this solution barely fails, but by the own admission of the authors the presented bound is too conservative. 
In fact, if one uses the appropriate volume of the orbifold for the analysis rather than the underlying torus, one finds that anisotropic compactifications can work, even though the parameter space of solutions is not too generous. 
This implies that there is an intermediate orbifold \ac{GUT} symmetry (see e.g.\ \cite{Quiros:2003gg} for a review). 
However, this also means that the smaller radii are of the order of the string scale, and as stressed in \cite{Witten:1996mz}, one must use \ac{CFT} (rather than classical geometry) to analyze the model. 
This is one of the reasons why this \WriteUp focuses on orbifold constructions.

\section{Spectrum}
\label{sec:Spectrum}

Given a compactification of the heterotic string, one can determine its spectrum, i.e.\ the properties of the massless and massive excitations. 
One usually proceeds in two steps, by first determining the spectrum ``after compactification'' and then the spectrum of deformation of the model in which certain \acp{VEV} get switched on. 
In this section we focus on the former. 

\subsection{Massless gauge fields}
\label{subsec:Massless_gauge_fields}

In general, only a subset of the $G_\text{het}$ gauge fields survive the orbifold projections. 
They can be determined by finding the roots $p\in\Lambda_{\mathfrak{g}_\text{het}}$, i.e.\ $p\cdot p=2$, which satisfy the projection conditions
\begin{equation}
 p\cdot V = p\cdot W=0\pmod1
\end{equation}
with the $p$ from \Cref{eq:p_gauge} for all shift $V$ and Wilson line $W$ vectors (cf.\ \Cref{sec:Strings_on_orbifolds_B}).

\subsection{Chiral zero modes}
\label{sec:Spectrum:Chiral_zero_modes}

The zero modes are solutions of the mass equation to vanishing mass. 
In all known examples they are chiral w.r.t.\ some, possibly discrete, symmetry. 
The computation of the massless spectrum, i.e.\ gauge and chiral zero modes, is straightforward though tedious if done by hand, and can conveniently be performed with dedicated tools such as the
\texttt{Orbifolder}~\cite{Nilles:2011aj}.

\begin{figure}[htb]
 \centering
 \includegraphics[width=0.8\linewidth]{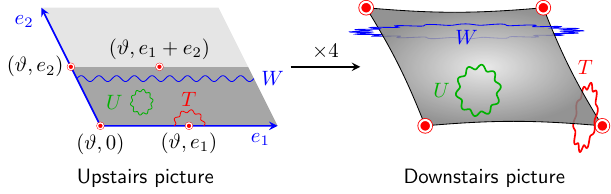}
 \caption{Cartoon of an orbifold. 
 Strings on orbifolds are in one-to-one correspondence with the conjugacy classes of the space group $\mathds{S}$. 
 Untwisted and twisted strings are associated with elements with trivial and nontrivial rotations, respectively. 
 In the depicted example of a $\mathds{T}^2/\mathds{Z}_2$ orbifold, the fundamental domain is half of the fundamental domain of the torus. 
 The edges can be identified (or ``glued together'') to obtain a pillow in the downstairs picture, where the twisted strings $T$ are localized at the corners. 
 Untwisted strings $U$ are free to propagate in the bulk, and winding strings $W$ wind around the torus.}
 \label{fig:Strings_on_orbifolds} 
\end{figure}

Explicit string models exhibit states beyond the \ac{SM} or \ac{MSSM} at some level. 
The additional states include the moduli as well as the winding and \ac{KK} modes, which we will review in \Cref{subsec:Moduli,subsec:Winding_and_KK_modes}, respectively. 
In addition, there are often vector-like states w.r.t.\ the \ac{SM} gauge group which are neither \ac{KK} nor winding modes. 
Whether or not these vector-like states are massless often depends on the point of moduli space under consideration. 
For instance, vector-like states may attain masses when giving \acp{VEV} to blow-up modes that smoothen out orbifold singularities, and break symmetries w.r.t.\ which these states are chiral. 
However, it would be arguably wrong to refer to the smoothened out version as ``cleaner'' since it is really the same model. 
In fact, often important properties of a given construction are much more directly accessible by studying the symmetry-enhanced point in field space, which is given by the orbifold point in this example, even though the vacuum is away from this point.

\subsection{Moduli}
\label{subsec:Moduli}

Virtually every supersymmetric string compactification contains fields which are classically flat directions, and this is in particular true for models that come close to the real world. 
Some of these so-called moduli do not have any charge under $G_\text{het}$, and comprise the K\"ahler moduli $\mathcal{T}_i$, the complex structure moduli $\mathcal{U}_i$ and the dilaton $\mathcal{S}$. 
Yet also some of the charged fields can attain \acp{VEV} because they are along $D$-flat directions. 
The \acp{VEV} of these fields determine, among other things, geometric properties of compact space. 
Classical flat directions can attain a nontrivial potential at the quantum level, in particular through nonperturbative effects. 
It is generally challenging to compute these potentials in full detail and thus determine the \acp{VEV} at the minima, see~\cite{McAllister:2023vgy} for more details of the analogous discussion in other versions of string theory.

\subsection{Winding and \ac{KK} modes}
\label{subsec:Winding_and_KK_modes}

The existence of winding and \ac{KK} modes is one of the most important features of string compactifications, and required to make the theory \ac{UV} complete. 
The properties of these modes are particularly accessible in torus-based compactifications such as toroidal orbifolds.

\section{Symmetries of the effective action}
\label{sec:Symmetries}

Heterotic compactifications lead to effective 4D theories exhibiting  various symmetries \cite{Dixon:1989fj}, which largely determine the phenomenological properties of the respective models. 
These symmetries include\begingroup\colorlet{blue}{black}
\begin{enumerate}
  \item either \href{subsec:SUSY}{$\mathcal{N}=1$ or no supersymmetry},
  \item \href{subsec:Continuous_gauge_symmetries}{continuous gauge symmetries},
  which mainly originate from the 10D $G_\text{het}$ symmetries, i.e.\ the root
  lattice of the 16 left-moving coordinates,
 \item \href{subsec:R-symmetries}{$R$ symmetries} (in \ac{SUSY} compactifications),\label{en:R}
 \item \href{subsec:Flavor_symmetries}{flavor symmetries}, 
 \item \href{subsec:Modular_symmetries}{modular symmetries}, and
 \item \href{subsec:Outer_automorphisms}{outer automorphisms} which may be \CP
  or \CP-like transformations.\footnote{It is important to distinguish between proper 
  \CP transformations, which map all particles to their own antiparticles, and 
  \CP-like transformations, which only send some of the particles to their 
  antiparticles~\cite{Chen:2014tpa}.}\label{en:outer} 
\end{enumerate}\endgroup
More recently, it has been pointed out that the symmetries~\ref{en:R}-\ref{en:outer} may be regarded as outer automorphisms of the Narain lattice~\cite{Baur:2019kwi,Baur:2019iai} in the Narain formulation of toroidal compactifications of the heterotic strings~\cite{Narain:1985jj} (see also~\cite{GrootNibbelink:2017usl}). 
The gauge symmetries can go beyond $G_\text{het}$ if the compact space has special properties, such as some radii equalling certain critical values (cf.\ e.g.\ \cite{GrootNibbelink:2017usl}).

\subsection{\ac{SUSY}}
\label{subsec:SUSY}

$\mathcal{N}=1$ \ac{SUSY} (cf.\ \Cref{subsec:BSM}) has long been a standard ingredient of string models. 
Whether or not $\mathcal{N}=1$ \ac{SUSY} is preserved by the compactifaction depends on the holonomy group of the compact space \cite{Candelas:1985en}. 
In the case of a smooth compactification, the requirement that the compactification preserves $\mathcal{N}=1$ \ac{SUSY} dictates that the manifold has to be of the \ac{CY} type, and in orbifolds it
requires the twist to fit into $\text{SU}(3)$, the holonomy group of \ac{CY} manifolds.

\subsection{Continuous gauge symmetries}
\label{subsec:Continuous_gauge_symmetries}

After compactification the residual continuous gauge symmetry, $G_\text{gauge}$, of a realistic model has to contain the gauge symmetry of the \ac{SM}~\eqref{eq:C_SM}. 
The gauge symmetry follows already from our discussion in \Cref{subsec:Massless_gauge_fields}. 
Apart from the obvious option that $G_\text{gauge}=G_\text{SM}\times G_\text{beyond}$ promising models may also replace $G_\text{SM}$ by the \ac{PS} group $G_\text{PS}=\mathrm{SU}(4)_\text{C}\times\mathrm{SU}(2)_\text{L}\times\mathrm{SU}(2)_\text{R}$, the so-called left-right symmetry \cite{Senjanovic:1975rk} $G_\text{LR}=\mathrm{SU}(3)_\text{C}\times\mathrm{SU}(2)_\text{L}\times\mathrm{SU}(2)_\text{R}\times\mathrm{U}(1)_{{B}-{L}}$, or the flipped $\mathrm{SU}(5)$ symmetry \cite{Barr:1981qv}. 
Other grand unified symmetries are in principle possible but may be challenged by doublet-triplet splitting problems and the lack of appropriate Higgs fields that break the larger symmetry to $G_\text{SM}$. 

Very often in geometric orbifolds with $\mathcal{N}=1$ \ac{SUSY} one $\text{U}(1)$ factor appears anomalous, with the anomaly being cancelled by the \ac{GS} mechanism. 
As a consequence, the $D$-term potential contains an \ac{FI} term, $\mathscr{V}_D\supset g^2 D_\text{anom}^2$, where
\begin{align}
  D_\text{anom} = \sum_i q_\text{anom}^i \left|\varphi_i\right|^2+\xi
  =0\quad  \text{with } \xi= \frac{g^2\,\operatorname{Tr}\mathsf{t}_\text{anom}}{192\pi^2}M_\text{P}^2\;.
  \label{eq:Danom}
\end{align}
$g$ denotes the gauge coupling, and $\mathsf{t}_\text{anom}$ the generator of $\text{U}(1)_\text{anom}$. 
The requirement of a vanishing of the $D$-term potential induces \acp{VEV} of $\text{U}(1)_\text{anom}$ charged fields $\varphi_i$ that breaks $\text{U}(1)_\text{anom}$ and in the overwhelming majority of cases further symmetries \cite{Font:1988tp}. 
Clearly, in realistic models the fields $\varphi_i$ acquiring large \acp{VEV} must be \ac{SM} singlets. 
Configurations with vanishing $D$-terms can be identified by constructing holomorphic monomials which are invariant under all gauge symmetries but carry nontrivial charge under $\text{U}(1)_\text{anom}$ \cite{Buccella:1982nx,Cleaver:1997nj,Cvetic:1998gv}. 
A complete basis of such monomials can be obtained via the Hilbert basis \cite{Kappl:2011vi}.  
In the vast majority of explicit models, $\operatorname{Tr}\mathsf{t}_\text{anom}\sim\mathcal{O}(100)$, so that $\sqrt{\xi}/M_\text{P}$ is of the order of the Cabibbo angle, and, therefore, the \acp{VEV} induced by \eqref{eq:Danom} may conceivably play a role in explaining flavor hierarchies~\cite{Binetruy:1994ru}. 

The extra symmetry $G_\text{beyond}$ is usually partly broken by the \acp{VEV} that cancel the \ac{FI} term. 
The residual continuous part can conceivably provide us with a hidden sector leading to dynamical \ac{SUSY} breakdown \cite{Derendinger:1985kk,Intriligator:2006dd}.
There are also usually discrete symmetries, which can be determined systematically with the Smith normal form \cite{Petersen:2009ip}.

\subsection{Discrete symmetries}

\subsubsection{Symmetries from a Narain compactification}
\label{subsec:Narain}

The Narain formulation provides an alternative to the usual toroidal
compactification  of the heterotic string discussed in \Cref{sec:Compactifications}. 
Let us consider first scenarios in which the six extra dimensions are compacitified in a $\mathds{T}^6$. 
In the Narain formulation, the $6$ right- and $6+16$ left-moving 
compact (bosonic) coordinates are considered independently, so that, taking into account also the gauge degrees of freedom, the $\mathds{T}^6$ compactification is specified in terms of an auxiliary $(2\cdot6+16)$D torus according to
\begin{equation}
 Y \sim Y + E \widehat{N}\;,\quad \text{where}\quad 
 Y = \left(X_R^i, X_L^i, X_L^I\right)^\mathrm{T}\;,\quad 
 \widehat{N}=(w,k,q)^\mathrm{T}\in\mathds{Z}^{2\cdot6+16}\;.
\end{equation}
Here, $E$ is the Narain vielbein which spans the $28$D even, integer and self-dual Narain lattice $\Gamma$ of signature $(6,6+16)$. Further, the integer vector $\widehat{N}$ includes the winding numbers $w^i$, the \ac{KK} numbers $k^i$, and the gauge momenta $q^I$ discussed in \Cref{subsec:Continuous_gauge_symmetries}. 
The properties of $\Gamma$ are encoded in the condition
\begin{equation}
  E^\mathrm{T}\eta E = \begin{pmatrix} 
  	0 & \mathds{1}_D & 0 \\ 
	\mathds{1}_D & 0 & 0 \\ 
	0 & 0 & g
  \end{pmatrix} =: \widehat{\eta}\;,
\end{equation}
where $\eta$ is the flat metric with signature $(6,6+16)$, $g$ denotes the 16D Cartan matrix of the heterotic string, and we have defined the Narain metric $\widehat{\eta}$.

It turns out that the group of outer automorphisms of $\Gamma$, defined by
\begin{equation}
  \text{O}_{\widehat{\eta}}(6,6+16,\mathds{Z}) ~=~ 
  \left\{{} \widehat{\Sigma} ~|~ \widehat{\Sigma}^\mathrm{T}
  \widehat{\eta}\,\widehat{\Sigma}\, {}\right\}\qquad \text{with}\quad 
  \widehat{\Sigma}\in\text{GL}(2\cdot6+16,\mathds{Z})\;,
\end{equation}
describes all the discrete symmetries of the toroidal compactification.
Hence, naturally  $\mathrm{O}_{\widehat{\eta}}(6,6+16,\mathds{Z})$ contains the modular transformations of the compactification, including mirror symmetries and \CP-like transformations of the moduli.

From this compactification, it is easy to arrive at the symmetries of a toroidal orbifold. 
In this formalism, an orbifold is obtained by modding out a subgroup of $\Gamma$. 
Let us consider, for simplicity, the case of an Abelian orbifold
without roto-translations. 
Treating left- and right-moving coordinates as independent, as before, the orbifold identification is given by
\begin{equation}
Y \sim \Theta^r Y + E \widehat{N}\;,
\end{equation}
with the Narain twist
\begin{equation}
  \Theta = 
  \operatorname{diag}\bigl( 
       \vartheta_\text{R},
       \vartheta_\text{L},
       \vartheta_g  
  \bigr)
  \;,
  \quad\text{where }\vartheta_\text{R},\vartheta_\text{L}\in\mathrm{O}(6)
  \text{ and } \vartheta_g\in\text{O}(16)\;.
\end{equation}
We can further impose the  orbifold to be of order $N$ by demanding $\Theta^N = \mathds{1}$. 
The Narain twist must leave the chosen Narain lattice invariant, i.e.\ $\Theta\Gamma = \Gamma$, which ensures that the moduli remain invariant under the orbifold action.
Hence, some of the moduli of the original toroidal compactification are hereby fixed. 
Note that the possibility $\vartheta_\text{L}\neq\vartheta_\text{R}$ defines an asymmetric orbifold. 
Limiting ourselves to $\vartheta_\text{L}=\vartheta_\text{R}=\vartheta$, we recover the geometric picture of the symmetric orbifolds introduced in \Cref{sec:Compactifications}.

The discrete symmetries of the orbifold include then the subgroup of rotational outer automorphisms of the toroidal compactification,
$\widehat{\Sigma}\in\text{O}_{\widehat{\eta}}(6,6+16,\mathds{Z})$, that are left unbroken by the orbifold, i.e.\ which satisfy
\begin{equation}
\label{eq:OrbifoldInvarianceNarainRotation}
 \widehat{\Sigma}^{-1}\widehat{\Theta}^k\widehat{\Sigma} = \widehat{\Theta}^{k'}\;, 
 \quad\text{where}\quad
 k,k'=1,\ldots,N\;,
\end{equation}
and $\widehat{\Theta} = E^{-1}\Theta\,E$ is the Narain twist in the Narain lattice basis.
In addition, now there are translational outer  automorphisms of the orbifold\footnote{The Narain twist combines with the  translations of the Narain lattice to build the Narain space group $\mathds{S}_\text{Narain}$. 
Formally, it is the outer automorphisms of $S_\mathrm{Narain}$ that we refer here as the automorphisms of the orbifold.} given by 
\begin{equation}
Y \sim Y + E \widehat{T}\;,\qquad\text{with}\qquad \widehat{T}\not\in\mathds{Z}^{2\cdot6+16}\,.
\end{equation}
In order to be compatible with the orbifold, the translations must fulfill
\begin{equation}
\label{eq:OrbifoldInvarianceNarainTranslation}
  \left(\mathds{1}_{2\cdot6+16}-\widehat\Theta^k\right)\,\widehat{T}\in\Gamma\,,\qquad 1\leq k\leq N\,.
\end{equation}
Note that these translations build a normal subgroup of the full group of outer automorphism of the orbifold.

These discrete residual transformations give rise to $R$, flavor, modular and outer automorphism symmetries, which we will discuss separately in what follows.

\subsubsection{$R$ symmetries}
\label{subsec:R-symmetries}

Supersymmetric orbifold compactifications usually do not break the Lorentz symmetry of the compact 6 dimensions completely but leave discrete remnants which act as $R$ symmetries in the effective description. 
Since the superpotential has a nontrivial modular weight, modular transformations, which we will discuss in more detail below in \Cref{subsec:Modular_symmetries}, are generically $R$ symmetries. 
As we shall see in an explicit example in \Cref{subsec:Example_with_Neq1_SUSY}, certain $R$ symmetries can be instrumental in resolving some of the phenomenological issues.

\subsubsection{Flavor symmetries}
\label{subsec:Flavor_symmetries}

The repetition of families in the \ac{SM} begs for an explanation. Flavor symmetries may address this question. 
String compactifications can give rise to non-Abelian discrete symmetries in which the three generations of the \ac{SM} transform as a $\boldsymbol{3}$-plet, or two generations as a $\boldsymbol{2}$-plet. 
Such symmetries may arguably play a role in understanding the flavor structure of the \ac{SM} (see e.g.\ \cite{Ishimori:2010au} for references).

In the geometric approach, flavor symmetries can be obtained from the replication of matter states at different yet equivalent orbifold singularities in the compact dimensions~\cite{Ploger:2007iq}. 
The emerging permutations combine with additional symmetries from the string selection rules to non-Abelian discrete symmetries.  
These rules act on matter fields as Abelian symmetries of the effective theory, which can be understood as an Abelianization of the space group of the orbifold~\cite{Ramos-Sanchez:2018edc}. 
It has been verified in explicit examples that the above-mentioned non-Abelian symmetries emerge from continuous gauge symmetries~\cite{Beye:2014nxa} are hence gauged, as one would expect. 

In the Narain formalism, these symmetries are identified with the subgroup of translational outer automorphisms of the orbifold.

\subsubsection{Modular symmetries}
\label{subsec:Modular_symmetries}

Modular symmetries are ubiquitous in string compactifications. 
They are symmetries of certain loop diagrams and the partition function. 
Moreover, toroidal orbifold compactifications exhibit modular symmetries. 
It is important to distinguish between the two. 

World-sheet modular invariance has far-reaching implications for the \ac{UV} consistency of the theory, the comprehensive discussion of which is beyond the scope of this \WriteUp. 
In particular, modular invariance conditions constrain the choices of the geometrical data of the models~\cite{Dixon:1986jc,Vafa:1986wx}. 
Among other things, they ensure that the models are free of anomalies. 

Target-space modular invariance provides us with important constraints on the K\"ahler potential and couplings of the theory~\cite{LopesCardoso:1994is}. 
These modular symmetries contain crucial information on the couplings of the theory \cite{Ferrara:1989bc}, and even provide us with an alternative to the CFT computation \cite{Dixon:1986qv}.

In the geometric approach of toroidal orbifold compactifications, the properties of target-space modular symmetries have been explored. 
Among other features, these symmetries are free of anomalies thanks to the \ac{GS} mechanism. 
Further, the transformation of matter fields under these symmetries have been determined. 
Denoting by\footnote{Here, we adopt the convention  $T =\frac{1}{\alpha'}\left(B+\ii \sqrt{\det G}\right)$, where $B$ is the nontrivial component of the antisymmetric $B$-field and $G$ the metric of the 2D orbifold sector, respectively.} $T$ the K\"ahler  modulus of a $\mathds{T}^2/\mathds{Z}_N$ orbifold sector, by $\gamma$ any transformation from the corresponding $\mathrm{SL}(2,\mathds{Z})$ modular group, the $p$\textsuperscript{th} multiplet $\Phi_p$ of twisted matter fields of the orbifold transform according to~\cite{Lauer:1989ax,Lauer:1990tm}
\begin{equation}
\label{eq:ModularTrafoOrbifold}
 \Phi_p \xmapsto{~\gamma~} (c T + d)^{n_p}\,\rho(\gamma)\,\Phi_p\;,\qquad 
 \gamma=\begin{pmatrix}a&b\\c&d\end{pmatrix}\in\text{SL}(2,\mathds{Z})\;.
\end{equation}
Here $\rho(\gamma)$ is a representation of $\gamma$ in a finite (double cover) modular group $\Gamma'_A = \text{SL}(2,\mathds{Z})/\Gamma(A)$ with $A$ depending on the  order $N$ of the orbifold, and $n_p$ is the (possibly fractional) modular weight carried by the twisted fields~\cite{Ibanez:1992hc}.

The 4D effective supersymmetric field theory of such a model is governed by these symmetries. In particular, the modular transformation of the associated K\"ahler potential reads at leading order
\begin{equation}
\label{eq:KaehlerOrbifold}
 K = -\ln\bigl(-\ii T+\ii\overline{T}\bigr) + 
 \sum_p \bigl(-\ii T+\ii \overline{T}\bigr)^{n_p}|\Phi_p|^2\;.
\end{equation}
This transformation is cancelled by a K\"ahler transformation of the superpotential provided that the superpotential terms of order $m$ are given by
\begin{equation}
\label{eq:SuperpotentialOrbifold}
 \mathscr{W} \supset Y(T)\,\Phi_{p_1} \cdots \Phi_{p_m}
\end{equation}
and have total modular weight $-1$ per complex orbifold plane, where $Y(T)$ is a modular form.

In the Narain formalism, the modular symmetries are identified with the subgroup of rotational outer automorphisms of the orbifold. 
Note that the $R$ symmetries (cf.\ \Cref{subsec:R-symmetries}) may also be understood as remnants of the target-space modular transformations of the complex structure moduli of the orbifold. 
This implies a relation between the $R$ charges of matter fields and their so-called modular weights~\cite{Nilles:2020gvu}.

\subsubsection{Outer automorphisms}
\label{subsec:Outer_automorphisms}

The effective action can exhibit certain outer-automorphism symmetries. 
These symmetries contain fundamental transformations like charge conjugation $\mathcal{C}$, parity $\mathcal{P}$ and time reversal $\mathcal{T}$. 
$\mathcal{CP}$ has to be broken in the flavor sector order to describe the real world, a criterion that already some of the first explicit string models turn out to satisfy \cite{Nilles:2018wex}. 
Further outer automorphisms comprise the left-right parity of the left-right symmetric model, which may emerge as discrete remnants of the continuous gauge symmetries after orbifolding \cite{Biermann:2019amx}.

Note that in the Narain formalism some outer automorphisms of the orbifold can also be considered $\mathcal{CP}$-like transformations. 
It remains to be seen whether there is a connection between such transformations and the physical $\mathcal{CP}$. 

\subsection{Approximate symmetries and hierarchies}
\label{subsec:approximateSym}

Starting at a symmetry-enhanced point has various benefits compared to analyzing generic points in moduli space. 
The models reviewed in this \WriteUp give rise to a variety of mildly broken, and thus approximate, symmetries. 
As already mentioned above, the latter may conceivably explain the observed hierarchies in the flavor sector \cite{Binetruy:1994ru}. 
They may also provide us with solutions to the $\mu$ and/or strong $\mathcal{CP}$ problems (cf.\ \cite{Kappl:2008ie,Choi:2009jt}). 
They may explain the scales in models of dynamical supersymmetry breaking (such as \cite{Intriligator:2006dd}, which requires an explicit mass for some pairs of vector-like states) or the messengers of gauge mediated \ac{SUSY} breaking \cite{Dine:1995ag}.

\section{Challenges}
\label{sec:Challenges}

Modern days model building faces various challenges. 
Bottom-up models usually can accommodate observation but the shear abundance and flexibility of the emerging constructions make it appear unlikely that these activities alone will provide us with unique answers. 
This \WriteUp focuses on top-down models, which come with their own challenges. 
They include:
\begin{challenges}
 \item Obtain the correct gauge symmetry.\label{Challenge:GaugeSymmetry}
 \item Obtain the correct spectrum, i.e.\ the three generations of quarks and
  leptons without any other states which are chiral w.r.t.\
  $G_\text{SM}$.\label{Challenge:Spectrum}  
 \item Avoid dangerous operators such as those leading to too fast proton decay or too large \acp{FCNC}.\label{Challenge:OffendingOperators}
 \item Provide a consistent cosmological history.\label{Challenge:Cosmology} 
 \item Reproduce the observed values of continuous parameters of the \ac{SM}, i.e.\ the gauge and Yukawa couplings.\label{Challenge:Couplings} 
\end{challenges}

The first challenge, \cref{Challenge:GaugeSymmetry}, has been mastered
successfully in heterotic model building early on. 
Compactification breaks the gauge symmetry of the 10D heterotic string, and it is fairly straightforward to obtain the \ac{SM} gauge symmetry, or a symmetry that can be broken to $G_\text{SM}$.

Obtaining the correct spectrum, i.e.\ \cref{Challenge:Spectrum}, has been a bigger challenge since string models may yield the wrong number of generations or give rise to chiral exotics. 
Nonetheless, extensive scans accompanied with appropriate search strategies has enabled the community to identify a large number of compactifications that exhibit the chiral spectrum of the \ac{SM} at low energies.\footnote{However, the number of chiral generations at low energies and at the compactification scale may be different \cite{Ramos-Sanchez:2021woq}, so some care needs to be taken not to prematurely discard models.} 
Notice that this often involves the appearance of extra, vector-like states which acquire masses below the compactification scale. 
While there are some constraints on such states, e.g.\ from the requirement that the gauge couplings remain perturbative, they also may play an important role in the phenomenology of the model, cf.\ our discussion in \Cref{sec:Spectrum:Chiral_zero_modes}. 
Another concern stems from so-called fractionally charged exotics. As already mentioned in \Cref{subsec:EarlyUniverse}, there are tight experimental constraints on their relic abundance (cf.\ e.g.\ \cite{Langacker:2011db,Halverson:2018vbo}), so the appearance of such states in the spectrum leads to the requirement that they are not produced copiously in the early universe.

The offending operators mentioned in \cref{Challenge:OffendingOperators} include the so-called $R$-parity violating couplings of the \ac{MSSM}, cf.\ our discussion below \Cref{eq:W_MSSM}. 
Forbidding this couplings requires additional symmetries, the simplest option being $R$- or matter parity. 
This shows, in particular, that it is not sufficient to obtain models with three generations and the \ac{SM} gauge symmetry, one necessarily needs additional symmetries. 
The offending operators also often get induced by extra states. 
For instance, $\text{SU}(3)$ triplet partners of the \ac{MSSM} Higgs doublets may mediate proton decay at an unacceptably large rate \cite{Sakai:1981pk}. 
This problem haunts 4D models of grand unification but is absent in certain higher-dimensional variants \cite{Altarelli:2001qj}.  
Nonetheless, it may get reintroduced through other vector-like states.

While string cosmology is an active field (see \cite{McAllister:2023vgy} for more details), only limited attention has been given to the performance of otherwise promising models, which
overcome \cref{Challenge:GaugeSymmetry,Challenge:Spectrum,Challenge:OffendingOperators,Challenge:OffendingOperators}.
It remains a task for the future to see whether, say, inflation can be realized and a realistic baryon asymmetry can be generated.

Challenge~\Cref{Challenge:Couplings} is major, and has not been completely mastered in any known construction so far, let alone in remotely realistic models. 
Part of the problem is that the couplings depend on the \acp{VEV} of certain scalar fields, the moduli (cf.\ \Cref{subsec:Moduli}) and possibly other fields.\footnote{The precise definition of what a modulus is varies over the literature. In some parts $D$-flat combinations of the other fields are referred to as moduli.} 
This means that mastering challenge~\cref{Challenge:Spectrum} requires stabilizing all moduli. 
This is a topic on its own, which is covered in \cite{McAllister:2023vgy}. 
Within the examples given in \Cref{sec:Examples}, we will comment on the extent to which realistic couplings are obtained.

\section{Examples}
\label{sec:Examples}

\subsection{Geometric orbifold with $\mathcal{N}=1$ \ac{SUSY}}
\label{subsec:Example_with_Neq1_SUSY}

Rather than reviewing extensive model scans, let us focus on a particular example, the model of \cite{Kappl:2010yu}. The
orbifold has noncontractible cycles (cf.\ \Cref{subsec:Classification_of_toroidal_orbifold_geometries}) which allow one to break an $\text{SU}(5)$ grand unified symmetry nonlocally down to $G_\text{SM}$. This type of \ac{GUT} breaking avoids fractionally charged exotics. The spectrum consists of 3 chiral generations of quark and leptons plus additional states which are vectorlike w.r.t.\ $G_\text{SM}$ but massless at the orbifold point, see \Cref{tab:DiscreteChargesMatterZ2xZ2_1-1}.  Like the majority of models of this type, there is an \ac{FI} term which has to be cancelled consistently with vanishing of the $F$- and (other) $D$-terms. The corresponding \acp{VEV} break the gauge symmetry at the orbifold point down to
\begin{equation}
 G_\text{residual}=G_\text{SM}\times\mathds{Z}_4^R\times\text{SU}(2)_\text{hid}\;,
\end{equation}
where $G_\text{SM}$ and $\text{SU}(2)_\text{hid}$ stem from two different $\text{E}_8$ factors, and none of the \ac{SM} matter is charged under $\text{SU}(2)_\text{hid}$.

\begin{table}[htb]
\centering
\begin{subtable}{0.3\linewidth}
\centering
\begin{tabular}{l*{5}{@{\,}c@{\,}}}
\toprule[0.8pt]
 & $q_i$ & $\bar u_i$ & $\bar d_i$ & $\ell_i$ & $\bar e_i$ \\
$\mathds{Z}_4^R$ & 1 & 1 & 1 & 1 & 1\\
\bottomrule[0.8pt]
\end{tabular}
\caption{Quarks and leptons.}
\end{subtable}\hspace{1ex}
\begin{subtable}{0.68\linewidth}
\centering
\begin{tabular}{l*{18}{@{\,}c@{\,}}}
\toprule[0.8pt]
  & $h_1$ & $h_2$& $h_3$& $h_4$& $h_5$& $h_6$ & $\bar h_1$ & $\bar h_2$ & $\bar h_3$ & $\bar h_4$ & $\bar h_5$ &  $\bar h_6$ & $\delta_1$ & $\delta_2$ & $\delta_3$ & $\bar\delta_1$ & $\bar\delta_2$ & $\bar\delta_3$ \\
 $\mathds{Z}_4^R$ & 0 & 2 & 0 & 2 & 0 & 0 & 0 & 2 & 0 & 0 & 2 & 2 & 0 & 2 & 2 & 2 & 0 & 0\\
 \bottomrule[0.8pt]
\end{tabular}
\caption{Higgs and exotics.}
\end{subtable}
\caption{$\mathds{Z}_4^R$ charges of the (a) matter fields and (b) Higgs and exotics. The
index $i$ in (a) takes values $i=1,2,3$. The $\ell_i$ and $h_i$ as well as the
$\bar d_i$ and $\bar\delta_i$ are distinguished by their $\mathds{Z}_4^R$
charges.}
\label{tab:DiscreteChargesMatterZ2xZ2_1-1}
\end{table}

These \acp{VEV} also provide mass terms for all \ac{SM} charged exotics, yet the $\mathds{Z}_4^R$ symmetry forbids the mass of one linear combination of Higgs fields, which gets identified with the \ac{MSSM} Higgs pair. This pair acquires a mass after $\mathds{Z}_4^R$ breaking. The order parameter of $R$ symmetry breaking is the gravitino mass, i.e.\ of the order of the soft terms, which are assumed to be not too far above the electroweak scale. That is, the $\mathds{Z}_4^R$, which is a discrete remnant of the Lorentz symmetry of compact space, can provide us with a solution to the $\mu$ problem along the lines of \cite{Antoniadis:1994hg}. In addition, this $\mathds{Z}_4^R$ suppresses dimension-5 proton decay operators enough to be consistent with observation.

It has been checked that qualitatively realistic fermion masses arise, i.e.\ the Yukawa couplings have full rank, exhibit hierarchies and lead to nontrivial flavor mixing, and neutrino masses are see-saw suppressed. However, this is not to say that they are fully realistic, cf.\ our discussion of \cref{Challenge:Couplings}.

Altogether this example shows that explicit string models can successfully address some of the most pressing questions of (traditional) unified model building, including the $\mu$ and proton decay problems. However, it also illustrates that there is still a long way to go before we can claim to have found ``the'' stringy \ac{SM}. Apart from the question whether or not low-energy \ac{SUSY} is realized in Nature, one has to successfully fix the moduli. While this is a topic on its own, which is covered in \cite{McAllister:2023vgy}, the $\mathds{Z}_4^R$ symmetry and charges can be used to show that generically there are no flat directions. States with odd $\mathds{Z}_4^R$ acquire masses because the mass terms carry $\mathds{Z}_4^R$ charge $2\pmod4$, and the fields of $\mathds{Z}_4^R$ charge 2 pair up with linear combinations of $\mathds{Z}_4^R$ charge 0 fields. Of course, generic statements do not always lead to the correct conclusions, and one has to verify explicitly that there are no flat directions, what the possible \acp{VEV} of the $\mathds{Z}_4^R$ charge 0 fields are, and whether they leads to phenomenologically viable couplings in the \ac{SM} sector.

\subsection{Geometric orbifolds without \ac{SUSY}}
\label{sec:nonSUSY}

There is a consistent nonsupersymmetric heterotic
string~\cite{Alvarez-Gaume:1986ghj,Dixon:1986jc,Dixon:1986iz}, which can be
understood as a freely acting $\mathds{Z}_2$ orbifold of a $\mathcal{N}=1$
heterotic string~\cite{Dixon:1986jc,Dixon:1986iz}. Given the absence of
evidence of supersymmetry at colliders, this version of the heterotic
string deserves increasing attention, even though it does not exhibit the
protection that \ac{SUSY} offers against the appearance of tachyons, quadratic
divergences and a large cosmological constant~\cite{GrootNibbelink:2017luf}.

The massless spectrum of this theory comprises three components: the gravitational part
includes the graviton, the antisymmetric 2-form $B_{MN}$ and the dilaton; the gauge
bosons of $\mathrm{SO}(16)\times\mathrm{SO}(16)$ arise in the gauge sector; and
the charged matter states build the representations
$(\boldsymbol{128},\boldsymbol1)+(\boldsymbol1,\boldsymbol{128})+(\boldsymbol{16},\boldsymbol{16})$.

Applying similar compactification techniques such as orbifolds as in the supersymmetric case, there
has been some effort to study the phenomenology of compactifications of this
string theory in 4D, including models with a tachyon-free \acp{GUT} or \ac{SM} massless
spectrum~\cite{Blaszczyk:2014qoa,Abel:2015oxa,Abel:2017vos,Perez-Martinez:2021zjj}.
Although the progress does not yet compare to the supersymmetric case,
some general features in \ac{SM}-like models are known. In particular, the following
properties of the massless spectrum are found:
\begin{enumerate*}[label=(\roman*)]
  \item at perturbative level, tachyons can be avoided;
  \item models with only one \ac{SM} Higgs exist, but most of them exhibit a
    larger number of Higgses;
  \item there appear many fermion and scalar exotic states although there are
    models with a very small exotic spectrum;
  \item among the exotics, there are $\mathcal{O}(100)$ right-handed neutrinos;
  \item leptoquark scalars are present in different amounts; and
  \item the number of fermions and bosons can coincide, yielding the possibility
    of an exponentially suppressed one-loop cosmological constant.
\end{enumerate*}

As an example, let us focus on the model 2 of~\cite{Perez-Martinez:2021zjj}, based on an
Abelian orbifold compactification of the $\mathcal{N}=0$ string (in the bosonic formulation).
It includes the \ac{SM} gauge group and additional $\text{SU}(2)$ and $\text{U}(1)$
factors. In the fermionic sector, there are only three \ac{SM} generations arising from twisted
sectors and 119 right-handed neutrinos. In the scalar sector, besides 9 Higgs doublets
and 9 scalar leptoquarks, there are 30 \ac{SM} singlets, which may be considered flavons
of a traditional $\Delta(54)$ flavor symmetry. The modular flavor features of this kind
of models are not known.

\section{Smooth compactifications}
\label{sec:Smooth_Compactifications}

As already mentioned, one can obtain smooth compactifications 
of the heterotic $\text{E}_8\times\text{E}_8$ theory in 10 dimensions. 
If the compactification is to preserve $\mathcal{N}=1$ supersymmetry in 
4 dimensions, the compact space has to be a \ac{CY} manifold~\cite{Candelas:1985en}. 
Models with the chiral spectrum of the \ac{MSSM} have been found in this approach, 
see e.g.~\cite{Bouchard:2005ag}. Notice that, a priori, it is not clear that 
every supergravity compactification of this type has a stringy origin 
\cite{GrootNibbelink:2015dvi}, but is expected that a substantial fraction 
of the models in the literature correspond to string models. Machine learning 
techniques have been utilized to efficiently find models with the gauge symmetry 
and chiral spectrum of the \ac{SM}~\cite{Ruehle:2020jrk}. It will be interesting 
to see if the absence of certain terms~\cite{Silverstein:1995re,Anderson:2022kgk} 
can be understood in terms of ordinary symmetries as is the case in the orbifold 
models discussed here, or if novel mechanisms are at play. In the latter case, 
this may provide us with new ways of overcoming \Cref{Challenge:OffendingOperators}. 
In passing, let us mention that a significant amount of smooth models can be obtained 
from orbifolds via blow-up (cf.\ e.g.~\cite{GrootNibbelink:2009wzz}). In particular, 
giving \acp{VEV} to fields that are massless at the orbifold point often amounts 
to resolving the orbifold singularities. A detailed discussion of these interesting 
topics is, however, beyond the scope of this \WriteUp.

\section{Where do we stand?}
\label{sec:HeteroticModels_Outlook}

The aim of heterotic model building is to reproduce and interpret particle physics in the heterotic string. 
This can be achieved by identifying appropriate compactifications. 
As we have discussed, various approaches have led to large sets of semi-realistic models that exhibit the matter spectrum of the standard model, its minimal supersymmetric version, as well as certain gauge extensions such as \acp{GUT}. 
Using various techniques, the effective symmetries of these constructions have been studied, which has yielded interesting implications for flavor physics, $\mathcal{CP}$ violation, proton stability, supersymmetry breaking, among other features.

However, a clear, let alone unique, picture has not yet emerged. 
The gauge and Yukawa couplings are, in principle, consistent with observation in a subset of the models. 
However, solid and precise predictions of the latter have remained largely elusive so far. 
This is hardly surprising. 
To see why, recall that we believe to know the Lagrange density of QCD in great detail but it remains a challenge to precisely compute basic quantities like the proton mass. 
In string phenomenology, the analogous analyses are even more challenging as the computation of many observables requires, among other things, a precise, quantitive understanding of moduli stabilization, which has not yet obtained. 
However, one can turn this around by saying that the explicit models provide us with a framework in which progress in these open questions can lead to testable predictions for the many parameters of the \ac{SM} as well as \ac{BSM} physics. 
We expect that this framework will also deliver a picture to address some of the pressing puzzles in cosmology.

\begin{acronym}
  \acro{BBN}{big bang nucleosynthesis}
  \acro{BSM}{beyond the standard model}
  \acro{CFT}{conformal field theory}
  \acro{CY}{Calabi--Yau}
  \acro{EFT}{effective field theory}
  \acro{FCNC}{flavor changing neutral current}
  \acro{FFF}{Free Fermionic Formulation}
  \acro{FI}{Fayet--Iliopoulos \cite{Fayet:1974jb}} 
  \acro{GGSO}{Generalized GSO Projections}
  \acro{GS}{Green--Schwarz \cite{Green:1984sg}}
  \acro{GGSO}{Generalized \ac{GSO}}
  \acro{GSO}{Gliozzi--Scherk--Olive \cite{Gliozzi:1976qd}}
  \acro{GUT}{Grand Unified Theory}
  \acro{KK}{Kaluza--Klein}
  \acro{LHC}{Large Hadron Collider}
  \acro{MSSM}{minimal supersymmetric standard model}
  \acro{PS}{Pati--Salam \cite{Pati:1974yy}} 
  \acro{QFT}{quantum field theory}
  \acro{SB}{symmetry based}
  \acro{SM}{standard model}
  \acro{SUSY}{supersymmetry}
  \acro{UV}{ultraviolet}
  \acro{VEV}{vacuum expectation value}
\end{acronym}

\enlargethispage{1cm}
\section*{Acknowledgments}

We would like to thank Carlo Angelantonj and Ignatios Antoniadis, and all the editors of the Handbook on Quantum Gravity, for inviting us to write this review. 
Big thanks go to our collaborators on this topic Wilfried Buchm\"uller, Mu-Chun Chen, Maximilian Fallbacher, Maximilian Fischer, Stefan Groot-Nibbelink, Koichi Hamaguchi, Rolf Kappl, V\'ictor Knapp-P\'erez, Oleg Lebedev, Xiang-Gan Liu, Hans Peter Nilles, Yessenia Olgu\'in-Trejo, Susha Parameswaran, Ricardo P\'erez-Mart\'inez, Felix Pl\"oger, Stuart Raby, Graham Ross, Fabian Ruehle, Andreas Trautner, Patrick Vaudrevange, and Ivonne Zavala.
The work of SRS is partially supported by UNAM-PAPIIT IN113223, CONACYT grant CB-2017-2018/A1-S-13051 and Marcos Moshinsky Foundation.
The work of MR is supported by National Science Foundation grants PHY-1915005 and PHY-2210283, and parts of the work by MR were performed at the Aspen Center for Physics, which is supported by National Science Foundation grant PHY-160761. The authors were also supported by  UC-MEXUS-CONACyT grant No.\ CN-20-38.

\bibliography{HeteroticOrbifoldModels_Literature}
\bibliographystyle{utphys}

\input{HeteroticOrbifoldModels_Acronyms}

\end{document}

%% file: HeteroticOrbifoldModels_Acronyms.tex
\begin{acronym}
  \acro{BBN}{big bang nucleosynthesis}
  \acro{BSM}{beyond the standard model}
  \acro{CFT}{conformal field theory}
  \acro{CY}{Calabi--Yau}
  \acro{EFT}{effective field theory}
  \acro{FCNC}{flavor changing neutral current}
  \acro{FFF}{Free Fermionic Formulation}
  \acro{FI}{Fayet--Iliopoulos \cite{Fayet:1974jb}} 
  \acro{GGSO}{Generalized GSO Projections}
  \acro{GS}{Green--Schwarz \cite{Green:1984sg}}
  \acro{GGSO}{Generalized \ac{GSO}}
  \acro{GSO}{Gliozzi--Scherk--Olive \cite{Gliozzi:1976qd}}
  \acro{GUT}{Grand Unified Theory}
  \acro{KK}{Kaluza--Klein}
  \acro{LHC}{Large Hadron Collider}
  \acro{MSSM}{minimal supersymmetric standard model}
  \acro{PS}{Pati--Salam \cite{Pati:1974yy}} 
  \acro{QFT}{quantum field theory}
  \acro{SB}{symmetry based}
  \acro{SM}{standard model}
  \acro{SUSY}{supersymmetry}
  \acro{UV}{ultraviolet}
  \acro{VEV}{vacuum expectation value}
\end{acronym}

%% file: HeteroticOrbifoldModelsArXiV.bbl
\providecommand{\href}[2]{#2}\begingroup\raggedright\begin{thebibliography}{10}

\bibitem{Pati:1974yy}
J.~C. Pati and A.~Salam, ``{Lepton Number as the Fourth Color},''
  \href{http://dx.doi.org/10.1103/PhysRevD.10.275}{{\em Phys. Rev. D}
  {\bfseries 10} (1974) 275--289}. [Erratum: Phys.Rev.D 11, 703--703 (1975)].

\bibitem{Fayet:1974jb}
P.~Fayet and J.~Iliopoulos, ``{Spontaneously Broken Supergauge Symmetries and
  Goldstone Spinors},''
  \href{http://dx.doi.org/10.1016/0370-2693(74)90310-4}{{\em Phys. Lett. B}
  {\bfseries 51} (1974) 461--464}.

\bibitem{Green:1984sg}
M.~B. Green and J.~H. Schwarz, ``{Anomaly Cancellation in Supersymmetric D=10
  Gauge Theory and Superstring Theory},''
  \href{http://dx.doi.org/10.1016/0370-2693(84)91565-X}{{\em Phys. Lett. B}
  {\bfseries 149} (1984) 117--122}.

\bibitem{Olive:1981tb}
D.~I. Olive, ``{Relations between grand unified and monopole theories},''.
  Invited talk given at Study Conf. on Unification of Fundamental Interactions
  II, Erice, Italy, Oct 6-14, 1981.

\bibitem{Candelas:1985en}
P.~Candelas, G.~T. Horowitz, A.~Strominger, and E.~Witten, ``{Vacuum
  Configurations for Superstrings},''
  \href{http://dx.doi.org/10.1016/0550-3213(85)90602-9}{{\em Nucl. Phys. B}
  {\bfseries 258} (1985) 46--74}.

\bibitem{Dixon:1985jw}
L.~J. Dixon, J.~A. Harvey, C.~Vafa, and E.~Witten, ``{Strings on Orbifolds},''
  \href{http://dx.doi.org/10.1016/0550-3213(85)90593-0}{{\em Nucl. Phys. B}
  {\bfseries 261} (1985) 678--686}.

\bibitem{Dixon:1986jc}
L.~J. Dixon, J.~A. Harvey, C.~Vafa, and E.~Witten, ``{Strings on Orbifolds.
  2.},'' \href{http://dx.doi.org/10.1016/0550-3213(86)90287-7}{{\em Nucl. Phys.
  B} {\bfseries 274} (1986) 285--314}.

\bibitem{Blaszczyk:2010db}
M.~Blaszczyk, S.~Groot~Nibbelink, F.~Ruehle, M.~Trapletti, and P.~K.~S.
  Vaudrevange, ``{Heterotic MSSM on a Resolved Orbifold},''
  \href{http://dx.doi.org/10.1007/JHEP09(2010)065}{{\em JHEP} {\bfseries 09}
  (2010) 065}, \href{http://arxiv.org/abs/1007.0203}{{\ttfamily arXiv:1007.0203
  [hep-th]}}.

\bibitem{Ibanez:2012zz}
L.~E. Ib{\'a}{\~n}ez and A.~M. Uranga, {\em {String theory and particle
  physics: An introduction to string phenomenology}}.
\newblock Cambridge University Press, 2, 2012.

\bibitem{Langacker:2017uah}
P.~Langacker, \href{http://dx.doi.org/10.1201/b22175}{{\em {The Standard Model
  and Beyond}}}.
\newblock Taylor \& Francis, 2017.

\bibitem{Baumann:2022mni}
D.~Baumann, \href{http://dx.doi.org/10.1017/9781108937092}{{\em {Cosmology}}}.
\newblock Cambridge University Press, 7, 2022.

\bibitem{Langacker:2011db}
P.~Langacker and G.~Steigman, ``{Requiem for an FCHAMP? Fractionally CHArged,
  Massive Particle},'' \href{http://dx.doi.org/10.1103/PhysRevD.84.065040}{{\em
  Phys. Rev. D} {\bfseries 84} (2011) 065040},
  \href{http://arxiv.org/abs/1107.3131}{{\ttfamily arXiv:1107.3131 [hep-ph]}}.

\bibitem{Halverson:2018vbo}
J.~Halverson and P.~Langacker, ``{TASI Lectures on Remnants from the String
  Landscape},'' \href{http://dx.doi.org/10.22323/1.305.0019}{{\em PoS}
  {\bfseries TASI2017} (2018) 019},
  \href{http://arxiv.org/abs/1801.03503}{{\ttfamily arXiv:1801.03503
  [hep-th]}}.

\bibitem{Martin:1997ns}
S.~P. Martin, ``{A Supersymmetry primer},''
  \href{http://dx.doi.org/10.1142/9789812839657_0001}{{\em Adv. Ser. Direct.
  High Energy Phys.} {\bfseries 18} (1998) 1--98},
  \href{http://arxiv.org/abs/hep-ph/9709356}{{\ttfamily arXiv:hep-ph/9709356}}.

\bibitem{Kim:1983dt}
J.~E. Kim and H.~P. Nilles, ``{The mu Problem and the Strong CP Problem},''
  \href{http://dx.doi.org/10.1016/0370-2693(84)91890-2}{{\em Phys. Lett. B}
  {\bfseries 138} (1984) 150--154}.

\bibitem{Giudice:1988yz}
G.~F. Giudice and A.~Masiero, ``{A Natural Solution to the mu Problem in
  Supergravity Theories},''
  \href{http://dx.doi.org/10.1016/0370-2693(88)91613-9}{{\em Phys. Lett. B}
  {\bfseries 206} (1988) 480--484}.

\bibitem{Dimopoulos:1981yj}
S.~Dimopoulos, S.~Raby, and F.~Wilczek, ``{Supersymmetry and the Scale of
  Unification},'' \href{http://dx.doi.org/10.1103/PhysRevD.24.1681}{{\em Phys.
  Rev. D} {\bfseries 24} (1981) 1681--1683}.

\bibitem{Raby:2017ucc}
S.~Raby, \href{http://dx.doi.org/10.1007/978-3-319-55255-2}{{\em
  {Supersymmetric Grand Unified Theories}: {From Quarks to Strings via SUSY
  GUTs}}}, vol.~939.
\newblock Springer, 2017.

\bibitem{Breit:1985ud}
J.~D. Breit, B.~A. Ovrut, and G.~C. Segre, ``{E(6) Symmetry Breaking in the
  Superstring Theory},''
  \href{http://dx.doi.org/10.1016/0370-2693(85)90734-8}{{\em Phys. Lett. B}
  {\bfseries 158} (1985) 33}.

\bibitem{HandbookChapter1arxiv}
C.~Angelantonj and I.~Florakis, ``Introduction to String Theory,''. To appear
  as chapter of the Handbook of Quantum Gravity.

\bibitem{Gross:1984dd}
D.~J. Gross, J.~A. Harvey, E.~J. Martinec, and R.~Rohm, ``{The Heterotic
  String},'' \href{http://dx.doi.org/10.1103/PhysRevLett.54.502}{{\em Phys.
  Rev. Lett.} {\bfseries 54} (1985) 502--505}.

\bibitem{Donagi:2008xy}
R.~Donagi and K.~Wendland, ``{On orbifolds and free fermion constructions},''
  \href{http://dx.doi.org/10.1016/j.geomphys.2009.04.004}{{\em J. Geom. Phys.}
  {\bfseries 59} (2009) 942--968},
  \href{http://arxiv.org/abs/0809.0330}{{\ttfamily arXiv:0809.0330 [hep-th]}}.

\bibitem{Athanasopoulos:2016aws}
P.~Athanasopoulos, A.~E. Faraggi, S.~Groot~Nibbelink, and V.~M. Mehta,
  ``{Heterotic free fermionic and symmetric toroidal orbifold models},''
  \href{http://dx.doi.org/10.1007/JHEP04(2016)038}{{\em JHEP} {\bfseries 04}
  (2016) 038}, \href{http://arxiv.org/abs/1602.03082}{{\ttfamily
  arXiv:1602.03082 [hep-th]}}.

\bibitem{Narain:1986qm}
K.~S. Narain, M.~H. Sarmadi, and C.~Vafa, ``{Asymmetric Orbifolds},''
  \href{http://dx.doi.org/10.1016/0550-3213(87)90228-8}{{\em Nucl. Phys. B}
  {\bfseries 288} (1987) 551}.

\bibitem{Ibanez:1987pj}
L.~E. Ib{\'a}{\~n}ez, J.~Mas, H.-P. Nilles, and F.~Quevedo, ``{Heterotic
  Strings in Symmetric and Asymmetric Orbifold Backgrounds},''
  \href{http://dx.doi.org/10.1016/0550-3213(88)90166-6}{{\em Nucl. Phys. B}
  {\bfseries 301} (1988) 157--196}.

\bibitem{Gepner:1987sm}
D.~Gepner, ``{New Conformal Field Theories Associated with Lie Algebras and
  their Partition Functions},''
  \href{http://dx.doi.org/10.1016/0550-3213(87)90176-3}{{\em Nucl. Phys. B}
  {\bfseries 290} (1987) 10--24}.

\bibitem{Gepner:1987vz}
D.~Gepner, ``{Exactly Solvable String Compactifications on Manifolds of SU(N)
  Holonomy},'' \href{http://dx.doi.org/10.1016/0370-2693(87)90938-5}{{\em Phys.
  Lett. B} {\bfseries 199} (1987) 380--388}.

\bibitem{Fischer:2012qj}
M.~Fischer, M.~Ratz, J.~Torrado, and P.~K.~S. Vaudrevange, ``{Classification of
  symmetric toroidal orbifolds},''
  \href{http://dx.doi.org/10.1007/JHEP01(2013)084}{{\em JHEP} {\bfseries 01}
  (2013) 084}, \href{http://arxiv.org/abs/1209.3906}{{\ttfamily arXiv:1209.3906
  [hep-th]}}.

\bibitem{Giedt:2001zw}
J.~Giedt, ``{Spectra in standard - like Z(3) orbifold models},''
  \href{http://dx.doi.org/10.1006/aphy.2002.6231}{{\em Annals Phys.} {\bfseries
  297} (2002) 67--126}, \href{http://arxiv.org/abs/hep-th/0108244}{{\ttfamily
  arXiv:hep-th/0108244}}.

\bibitem{Ploger:2007iq}
F.~Pl{\"o}ger, S.~Ramos-S{\'a}nchez, M.~Ratz, and P.~K.~S. Vaudrevange,
  ``{Mirage Torsion},''
  \href{http://dx.doi.org/10.1088/1126-6708/2007/04/063}{{\em JHEP} {\bfseries
  04} (2007) 063}, \href{http://arxiv.org/abs/hep-th/0702176}{{\ttfamily
  arXiv:hep-th/0702176}}.

\bibitem{Vafa:1986wx}
C.~Vafa, ``{Modular Invariance and Discrete Torsion on Orbifolds},''
  \href{http://dx.doi.org/10.1016/0550-3213(86)90379-2}{{\em Nucl. Phys. B}
  {\bfseries 273} (1986) 592--606}.

\bibitem{Bailin:1999nk}
D.~Bailin and A.~Love, ``{Orbifold compactifications of string theory},''
  \href{http://dx.doi.org/10.1016/S0370-1573(98)00126-4}{{\em Phys. Rept.}
  {\bfseries 315} (1999) 285--408}.

\bibitem{Fischer:2013qza}
M.~Fischer, S.~Ramos-S{\'a}nchez, and P.~K.~S. Vaudrevange, ``{Heterotic
  non-Abelian orbifolds},''
  \href{http://dx.doi.org/10.1007/JHEP07(2013)080}{{\em JHEP} {\bfseries 07}
  (2013) 080}, \href{http://arxiv.org/abs/1304.7742}{{\ttfamily arXiv:1304.7742
  [hep-th]}}.

\bibitem{Hebecker:2004ce}
A.~Hebecker and M.~Trapletti, ``{Gauge unification in highly anisotropic string
  compactifications},''
  \href{http://dx.doi.org/10.1016/j.nuclphysb.2005.02.008}{{\em Nucl. Phys. B}
  {\bfseries 713} (2005) 173--203},
  \href{http://arxiv.org/abs/hep-th/0411131}{{\ttfamily arXiv:hep-th/0411131}}.

\bibitem{Bouchard:2005ag}
V.~Bouchard and R.~Donagi, ``{An SU(5) heterotic standard model},''
  \href{http://dx.doi.org/10.1016/j.physletb.2005.12.042}{{\em Phys. Lett. B}
  {\bfseries 633} (2006) 783--791},
  \href{http://arxiv.org/abs/hep-th/0512149}{{\ttfamily arXiv:hep-th/0512149}}.

\bibitem{Konopka:2012gy}
S.~J.~H. Konopka, ``{Non Abelian orbifold compactifications of the heterotic
  string},'' \href{http://dx.doi.org/10.1007/JHEP07(2013)023}{{\em JHEP}
  {\bfseries 07} (2013) 023}, \href{http://arxiv.org/abs/1210.5040}{{\ttfamily
  arXiv:1210.5040 [hep-th]}}.

\bibitem{Nilles:2014owa}
H.~P. Nilles and P.~K.~S. Vaudrevange, ``{Geography of Fields in Extra
  Dimensions: String Theory Lessons for Particle Physics},''
  \href{http://dx.doi.org/10.1142/S0217732315300086}{{\em Mod. Phys. Lett. A}
  {\bfseries 30} no.~10, (2015) 1530008},
  \href{http://arxiv.org/abs/1403.1597}{{\ttfamily arXiv:1403.1597 [hep-th]}}.

\bibitem{Olguin-Trejo:2018wpw}
Y.~Olgu{\'i}n-Trejo, R.~P{\'e}rez-Mart{\'i}nez, and S.~Ramos-S{\'a}nchez,
  ``{Charting the flavor landscape of MSSM-like Abelian heterotic orbifolds},''
  \href{http://dx.doi.org/10.1103/PhysRevD.98.106020}{{\em Phys. Rev. D}
  {\bfseries 98} no.~10, (2018) 106020},
  \href{http://arxiv.org/abs/1808.06622}{{\ttfamily arXiv:1808.06622
  [hep-th]}}.

\bibitem{Parr:2019bta}
E.~Parr and P.~K.~S. Vaudrevange, ``{Contrast data mining for the MSSM from
  strings},'' \href{http://dx.doi.org/10.1016/j.nuclphysb.2020.114922}{{\em
  Nucl. Phys. B} {\bfseries 952} (2020) 114922},
  \href{http://arxiv.org/abs/1910.13473}{{\ttfamily arXiv:1910.13473
  [hep-th]}}.

\bibitem{Witten:1996mz}
E.~Witten, ``{Strong coupling expansion of Calabi-Yau compactification},''
  \href{http://dx.doi.org/10.1016/0550-3213(96)00190-3}{{\em Nucl. Phys. B}
  {\bfseries 471} (1996) 135--158},
  \href{http://arxiv.org/abs/hep-th/9602070}{{\ttfamily arXiv:hep-th/9602070}}.

\bibitem{Dienes:1996du}
K.~R. Dienes, ``{String theory and the path to unification: A Review of recent
  developments},'' \href{http://dx.doi.org/10.1016/S0370-1573(97)00009-4}{{\em
  Phys. Rept.} {\bfseries 287} (1997) 447--525},
  \href{http://arxiv.org/abs/hep-th/9602045}{{\ttfamily arXiv:hep-th/9602045}}.

\bibitem{Quiros:2003gg}
M.~Quiros, ``{New ideas in symmetry breaking},'' in {\em {Theoretical Advanced
  Study Institute in Elementary Particle Physics (TASI 2002): Particle Physics
  and Cosmology: The Quest for Physics Beyond the Standard Model(s)}},
  pp.~549--601.
\newblock 2, 2003.
\newblock \href{http://arxiv.org/abs/hep-ph/0302189}{{\ttfamily
  arXiv:hep-ph/0302189}}.

\bibitem{Nilles:2011aj}
H.~P. Nilles, S.~Ramos-S{\'a}nchez, P.~K.~S. Vaudrevange, and A.~Wingerter,
  ``{The Orbifolder: A Tool to study the Low Energy Effective Theory of
  Heterotic Orbifolds},''
  \href{http://dx.doi.org/10.1016/j.cpc.2012.01.026}{{\em Comput. Phys.
  Commun.} {\bfseries 183} (2012) 1363--1380},
  \href{http://arxiv.org/abs/1110.5229}{{\ttfamily arXiv:1110.5229 [hep-th]}}.

\bibitem{McAllister:2023vgy}
L.~McAllister and F.~Quevedo, ``{Moduli Stabilization in String Theory},''
  \href{http://arxiv.org/abs/2310.20559}{{\ttfamily arXiv:2310.20559
  [hep-th]}}. To appear as chapter of the Handbook of Quantum Gravity.

\bibitem{Dixon:1989fj}
L.~J. Dixon, V.~Kaplunovsky, and J.~Louis, ``{On Effective Field Theories
  Describing (2,2) Vacua of the Heterotic String},''
  \href{http://dx.doi.org/10.1016/0550-3213(90)90057-K}{{\em Nucl. Phys. B}
  {\bfseries 329} (1990) 27--82}.

\bibitem{Chen:2014tpa}
M.-C. Chen, M.~Fallbacher, K.~T. Mahanthappa, M.~Ratz, and A.~Trautner, ``{CP
  Violation from Finite Groups},''
  \href{http://dx.doi.org/10.1016/j.nuclphysb.2014.03.023}{{\em Nucl. Phys. B}
  {\bfseries 883} (2014) 267--305},
  \href{http://arxiv.org/abs/1402.0507}{{\ttfamily arXiv:1402.0507 [hep-ph]}}.

\bibitem{Baur:2019kwi}
A.~Baur, H.~P. Nilles, A.~Trautner, and P.~K.~S. Vaudrevange, ``{Unification of
  Flavor, CP, and Modular Symmetries},''
  \href{http://dx.doi.org/10.1016/j.physletb.2019.03.066}{{\em Phys. Lett. B}
  {\bfseries 795} (2019) 7--14},
  \href{http://arxiv.org/abs/1901.03251}{{\ttfamily arXiv:1901.03251
  [hep-th]}}.

\bibitem{Baur:2019iai}
A.~Baur, H.~P. Nilles, A.~Trautner, and P.~K.~S. Vaudrevange, ``{A String
  Theory of Flavor and $\mathscr {CP}$},''
  \href{http://dx.doi.org/10.1016/j.nuclphysb.2019.114737}{{\em Nucl. Phys. B}
  {\bfseries 947} (2019) 114737},
  \href{http://arxiv.org/abs/1908.00805}{{\ttfamily arXiv:1908.00805
  [hep-th]}}.

\bibitem{Narain:1985jj}
K.~S. Narain, ``{New Heterotic String Theories in Uncompactified Dimensions
  \ensuremath{<} 10},''
  \href{http://dx.doi.org/10.1016/0370-2693(86)90682-9}{{\em Phys. Lett. B}
  {\bfseries 169} (1986) 41--46}.

\bibitem{GrootNibbelink:2017usl}
S.~Groot~Nibbelink and P.~K.~S. Vaudrevange, ``{T-duality orbifolds of
  heterotic Narain compactifications},''
  \href{http://dx.doi.org/10.1007/JHEP04(2017)030}{{\em JHEP} {\bfseries 04}
  (2017) 030}, \href{http://arxiv.org/abs/1703.05323}{{\ttfamily
  arXiv:1703.05323 [hep-th]}}.

\bibitem{Senjanovic:1975rk}
G.~Senjanovic and R.~N. Mohapatra, ``{Exact Left-Right Symmetry and Spontaneous
  Violation of Parity},''
  \href{http://dx.doi.org/10.1103/PhysRevD.12.1502}{{\em Phys. Rev. D}
  {\bfseries 12} (1975) 1502}.

\bibitem{Barr:1981qv}
S.~M. Barr, ``{A New Symmetry Breaking Pattern for SO(10) and Proton Decay},''
  \href{http://dx.doi.org/10.1016/0370-2693(82)90966-2}{{\em Phys. Lett. B}
  {\bfseries 112} (1982) 219--222}.

\bibitem{Font:1988tp}
A.~Font, L.~E. Ibanez, H.~P. Nilles, and F.~Quevedo, ``{Degenerate
  Orbifolds},'' \href{http://dx.doi.org/10.1016/0550-3213(88)90524-X}{{\em
  Nucl. Phys. B} {\bfseries 307} (1988) 109--129}. [Erratum: Nucl.Phys.B 310,
  764--764 (1988)].

\bibitem{Buccella:1982nx}
F.~Buccella, J.~P. Derendinger, S.~Ferrara, and C.~A. Savoy, ``{Patterns of
  Symmetry Breaking in Supersymmetric Gauge Theories},''
{\em Phys. Lett.} {\bfseries B115} (1982) 375.

\bibitem{Cleaver:1997nj}
G.~Cleaver, M.~Cvetic, J.~R. Espinosa, L.~L. Everett, and P.~Langacker,
  ``{Intermediate scales, mu parameter, and fermion masses from string
  models},'' \href{http://dx.doi.org/10.1103/PhysRevD.57.2701}{{\em Phys. Rev.
  D} {\bfseries 57} (1998) 2701--2715},
  \href{http://arxiv.org/abs/hep-ph/9705391}{{\ttfamily arXiv:hep-ph/9705391}}.

\bibitem{Cvetic:1998gv}
M.~Cvetic, L.~L. Everett, and J.~Wang, ``{Units and numerical values of the
  effective couplings in perturbative heterotic string vacua},''
  \href{http://dx.doi.org/10.1103/PhysRevD.59.107901}{{\em Phys. Rev. D}
  {\bfseries 59} (1999) 107901},
  \href{http://arxiv.org/abs/hep-ph/9808321}{{\ttfamily arXiv:hep-ph/9808321}}.

\bibitem{Kappl:2011vi}
R.~Kappl, M.~Ratz, and C.~Staudt, ``{The Hilbert basis method for D-flat
  directions and the superpotential},''
  \href{http://dx.doi.org/10.1007/JHEP10(2011)027}{{\em JHEP} {\bfseries 10}
  (2011) 027}, \href{http://arxiv.org/abs/1108.2154}{{\ttfamily arXiv:1108.2154
  [hep-th]}}.

\bibitem{Binetruy:1994ru}
P.~Binetruy and P.~Ramond, ``{Yukawa textures and anomalies},''
  \href{http://dx.doi.org/10.1016/0370-2693(95)00297-X}{{\em Phys. Lett. B}
  {\bfseries 350} (1995) 49--57},
  \href{http://arxiv.org/abs/hep-ph/9412385}{{\ttfamily arXiv:hep-ph/9412385}}.

\bibitem{Derendinger:1985kk}
J.~P. Derendinger, L.~E. Ib{\'a}{\~n}ez, and H.~P. Nilles, ``{On the Low-Energy
  d = 4, N=1 Supergravity Theory Extracted from the d = 10, N=1 Superstring},''
  \href{http://dx.doi.org/10.1016/0370-2693(85)91033-0}{{\em Phys. Lett. B}
  {\bfseries 155} (1985) 65--70}.

\bibitem{Intriligator:2006dd}
K.~A. Intriligator, N.~Seiberg, and D.~Shih, ``{Dynamical SUSY breaking in
  meta-stable vacua},''
  \href{http://dx.doi.org/10.1088/1126-6708/2006/04/021}{{\em JHEP} {\bfseries
  04} (2006) 021}, \href{http://arxiv.org/abs/hep-th/0602239}{{\ttfamily
  arXiv:hep-th/0602239}}.

\bibitem{Petersen:2009ip}
B.~Petersen, M.~Ratz, and R.~Schieren, ``{Patterns of remnant discrete
  symmetries},'' \href{http://dx.doi.org/10.1088/1126-6708/2009/08/111}{{\em
  JHEP} {\bfseries 08} (2009) 111},
  \href{http://arxiv.org/abs/0907.4049}{{\ttfamily arXiv:0907.4049 [hep-ph]}}.

\bibitem{Ishimori:2010au}
H.~Ishimori, T.~Kobayashi, H.~Ohki, Y.~Shimizu, H.~Okada, and M.~Tanimoto,
  ``{Non-Abelian Discrete Symmetries in Particle Physics},''
  \href{http://dx.doi.org/10.1143/PTPS.183.1}{{\em Prog. Theor. Phys. Suppl.}
  {\bfseries 183} (2010) 1--163},
  \href{http://arxiv.org/abs/1003.3552}{{\ttfamily arXiv:1003.3552 [hep-th]}}.

\bibitem{Ramos-Sanchez:2018edc}
S.~Ramos-S{\'a}nchez and P.~K.~S. Vaudrevange, ``{Note on the space group
  selection rule for closed strings on orbifolds},''
  \href{http://dx.doi.org/10.1007/JHEP01(2019)055}{{\em JHEP} {\bfseries 01}
  (2019) 055}, \href{http://arxiv.org/abs/1811.00580}{{\ttfamily
  arXiv:1811.00580 [hep-th]}}.

\bibitem{Beye:2014nxa}
F.~Beye, T.~Kobayashi, and S.~Kuwakino, ``{Gauge Origin of Discrete Flavor
  Symmetries in Heterotic Orbifolds},''
  \href{http://dx.doi.org/10.1016/j.physletb.2014.07.058}{{\em Phys. Lett. B}
  {\bfseries 736} (2014) 433--437},
  \href{http://arxiv.org/abs/1406.4660}{{\ttfamily arXiv:1406.4660 [hep-th]}}.

\bibitem{LopesCardoso:1994is}
G.~Lopes~Cardoso, D.~L{\"u}st, and T.~Mohaupt, ``{Moduli spaces and target
  space duality symmetries in (0,2) Z(N) orbifold theories with continuous
  Wilson lines},'' \href{http://dx.doi.org/10.1016/0550-3213(94)90594-0}{{\em
  Nucl. Phys. B} {\bfseries 432} (1994) 68--108},
  \href{http://arxiv.org/abs/hep-th/9405002}{{\ttfamily arXiv:hep-th/9405002}}.

\bibitem{Ferrara:1989bc}
S.~Ferrara, D.~L{\"u}st, A.~D. Shapere, and S.~Theisen, ``{Modular Invariance
  in Supersymmetric Field Theories},''
  \href{http://dx.doi.org/10.1016/0370-2693(89)90583-2}{{\em Phys. Lett. B}
  {\bfseries 225} (1989) 363}.

\bibitem{Dixon:1986qv}
L.~J. Dixon, D.~Friedan, E.~J. Martinec, and S.~H. Shenker, ``{The Conformal
  Field Theory of Orbifolds},''
  \href{http://dx.doi.org/10.1016/0550-3213(87)90676-6}{{\em Nucl. Phys. B}
  {\bfseries 282} (1987) 13--73}.

\bibitem{Lauer:1989ax}
J.~Lauer, J.~Mas, and H.~P. Nilles, ``{Duality and the Role of Nonperturbative
  Effects on the World Sheet},''
  \href{http://dx.doi.org/10.1016/0370-2693(89)91190-8}{{\em Phys. Lett. B}
  {\bfseries 226} (1989) 251--256}.

\bibitem{Lauer:1990tm}
J.~Lauer, J.~Mas, and H.~P. Nilles, ``{Twisted sector representations of
  discrete background symmetries for two-dimensional orbifolds},''
  \href{http://dx.doi.org/10.1016/0550-3213(91)90095-F}{{\em Nucl. Phys. B}
  {\bfseries 351} (1991) 353--424}.

\bibitem{Ibanez:1992hc}
L.~E. Ib{\'a\~n}ez and D.~L{\"u}st, ``{Duality anomaly cancellation, minimal
  string unification and the effective low-energy Lagrangian of 4-D strings},''
  \href{http://dx.doi.org/10.1016/0550-3213(92)90189-I}{{\em Nucl. Phys. B}
  {\bfseries 382} (1992) 305--361},
  \href{http://arxiv.org/abs/hep-th/9202046}{{\ttfamily arXiv:hep-th/9202046}}.

\bibitem{Nilles:2020gvu}
H.~P. Nilles, S.~Ramos\textendash{}S{\'a}nchez, and P.~K.~S. Vaudrevange,
  ``{Eclectic flavor scheme from ten-dimensional string theory - II detailed
  technical analysis},''
  \href{http://dx.doi.org/10.1016/j.nuclphysb.2021.115367}{{\em Nucl. Phys. B}
  {\bfseries 966} (2021) 115367},
  \href{http://arxiv.org/abs/2010.13798}{{\ttfamily arXiv:2010.13798
  [hep-th]}}.

\bibitem{Nilles:2018wex}
H.~P. Nilles, M.~Ratz, A.~Trautner, and P.~K.~S. Vaudrevange, ``{$\mathcal{CP}$
  violation from string theory},''
  \href{http://dx.doi.org/10.1016/j.physletb.2018.09.053}{{\em Phys. Lett. B}
  {\bfseries 786} (2018) 283--287},
  \href{http://arxiv.org/abs/1808.07060}{{\ttfamily arXiv:1808.07060
  [hep-th]}}.

\bibitem{Biermann:2019amx}
S.~Biermann, A.~M\"utter, E.~Parr, M.~Ratz, and P.~K.~S. Vaudrevange,
  ``{Discrete remnants of orbifolding},''
  \href{http://dx.doi.org/10.1103/PhysRevD.100.066030}{{\em Phys. Rev. D}
  {\bfseries 100} no.~6, (2019) 066030},
  \href{http://arxiv.org/abs/1906.10276}{{\ttfamily arXiv:1906.10276
  [hep-ph]}}.

\bibitem{Kappl:2008ie}
R.~Kappl, H.~P. Nilles, S.~Ramos-S{\'a}nchez, M.~Ratz, K.~Schmidt-Hoberg, and
  P.~K.~S. Vaudrevange, ``{Large hierarchies from approximate R symmetries},''
  \href{http://dx.doi.org/10.1103/PhysRevLett.102.121602}{{\em Phys. Rev.
  Lett.} {\bfseries 102} (2009) 121602},
  \href{http://arxiv.org/abs/0812.2120}{{\ttfamily arXiv:0812.2120 [hep-th]}}.

\bibitem{Choi:2009jt}
K.-S. Choi, H.~P. Nilles, S.~Ramos-S{\'a}nchez, and P.~K.~S. Vaudrevange,
  ``{Accions},'' \href{http://dx.doi.org/10.1016/j.physletb.2009.04.028}{{\em
  Phys. Lett. B} {\bfseries 675} (2009) 381--386},
  \href{http://arxiv.org/abs/0902.3070}{{\ttfamily arXiv:0902.3070 [hep-th]}}.

\bibitem{Dine:1995ag}
M.~Dine, A.~E. Nelson, Y.~Nir, and Y.~Shirman, ``{New tools for low-energy
  dynamical supersymmetry breaking},''
  \href{http://dx.doi.org/10.1103/PhysRevD.53.2658}{{\em Phys. Rev. D}
  {\bfseries 53} (1996) 2658--2669},
  \href{http://arxiv.org/abs/hep-ph/9507378}{{\ttfamily arXiv:hep-ph/9507378}}.

\bibitem{Ramos-Sanchez:2021woq}
S.~Ramos-S{\'a}nchez, M.~Ratz, Y.~Shirman, S.~Shukla, and M.~Waterbury,
  ``{Generation flow in field theory and strings},''
  \href{http://dx.doi.org/10.1007/JHEP10(2021)144}{{\em JHEP} {\bfseries 10}
  (2021) 144}, \href{http://arxiv.org/abs/2109.01681}{{\ttfamily
  arXiv:2109.01681 [hep-th]}}.

\bibitem{Sakai:1981pk}
N.~Sakai and T.~Yanagida, ``{Proton Decay in a Class of Supersymmetric Grand
  Unified Models},'' \href{http://dx.doi.org/10.1016/0550-3213(82)90457-6}{{\em
  Nucl. Phys. B} {\bfseries 197} (1982) 533}.

\bibitem{Altarelli:2001qj}
G.~Altarelli and F.~Feruglio, ``{SU(5) grand unification in extra dimensions
  and proton decay},''
  \href{http://dx.doi.org/10.1016/S0370-2693(01)00650-5}{{\em Phys. Lett. B}
  {\bfseries 511} (2001) 257--264},
  \href{http://arxiv.org/abs/hep-ph/0102301}{{\ttfamily arXiv:hep-ph/0102301}}.

\bibitem{Kappl:2010yu}
R.~Kappl, B.~Petersen, S.~Raby, M.~Ratz, R.~Schieren, and P.~K.~S. Vaudrevange,
  ``{String-Derived MSSM Vacua with Residual R Symmetries},''
  \href{http://dx.doi.org/10.1016/j.nuclphysb.2011.01.032}{{\em Nucl. Phys. B}
  {\bfseries 847} (2011) 325--349},
  \href{http://arxiv.org/abs/1012.4574}{{\ttfamily arXiv:1012.4574 [hep-th]}}.

\bibitem{Antoniadis:1994hg}
I.~Antoniadis, E.~Gava, K.~S. Narain, and T.~R. Taylor, ``{Effective mu term in
  superstring theory},''
  \href{http://dx.doi.org/10.1016/0550-3213(94)90599-1}{{\em Nucl. Phys. B}
  {\bfseries 432} (1994) 187--204},
  \href{http://arxiv.org/abs/hep-th/9405024}{{\ttfamily arXiv:hep-th/9405024}}.

\bibitem{Alvarez-Gaume:1986ghj}
L.~Alvarez-Gaum{\'e}, P.~H. Ginsparg, G.~W. Moore, and C.~Vafa, ``{An O(16) x
  O(16) Heterotic String},''
  \href{http://dx.doi.org/10.1016/0370-2693(86)91524-8}{{\em Phys. Lett. B}
  {\bfseries 171} (1986) 155--162}.

\bibitem{Dixon:1986iz}
L.~J. Dixon and J.~A. Harvey, ``{String Theories in Ten-Dimensions Without
  Space-Time Supersymmetry},''
  \href{http://dx.doi.org/10.1016/0550-3213(86)90619-X}{{\em Nucl. Phys. B}
  {\bfseries 274} (1986) 93--105}.

\bibitem{GrootNibbelink:2017luf}
S.~Groot~Nibbelink, O.~Loukas, A.~M\"utter, E.~Parr, and P.~K.~S. Vaudrevange,
  ``{Tension Between a Vanishing Cosmological Constant and Non-Supersymmetric
  Heterotic Orbifolds},'' \href{http://dx.doi.org/10.1002/prop.202000044}{{\em
  Fortsch. Phys.} {\bfseries 68} no.~7, (2020) 2000044},
  \href{http://arxiv.org/abs/1710.09237}{{\ttfamily arXiv:1710.09237
  [hep-th]}}.

\bibitem{Blaszczyk:2014qoa}
M.~Blaszczyk, S.~Groot~Nibbelink, O.~Loukas, and S.~Ramos-S{\'a}nchez,
  ``{Non-supersymmetric heterotic model building},''
  \href{http://dx.doi.org/10.1007/JHEP10(2014)119}{{\em JHEP} {\bfseries 10}
  (2014) 119}, \href{http://arxiv.org/abs/1407.6362}{{\ttfamily arXiv:1407.6362
  [hep-th]}}.

\bibitem{Abel:2015oxa}
S.~Abel, K.~R. Dienes, and E.~Mavroudi, ``{Towards a nonsupersymmetric string
  phenomenology},'' \href{http://dx.doi.org/10.1103/PhysRevD.91.126014}{{\em
  Phys. Rev. D} {\bfseries 91} no.~12, (2015) 126014},
  \href{http://arxiv.org/abs/1502.03087}{{\ttfamily arXiv:1502.03087
  [hep-th]}}.

\bibitem{Abel:2017vos}
S.~Abel, K.~R. Dienes, and E.~Mavroudi, ``{GUT precursors and entwined SUSY:
  The phenomenology of stable nonsupersymmetric strings},''
  \href{http://dx.doi.org/10.1103/PhysRevD.97.126017}{{\em Phys. Rev. D}
  {\bfseries 97} no.~12, (2018) 126017},
  \href{http://arxiv.org/abs/1712.06894}{{\ttfamily arXiv:1712.06894
  [hep-ph]}}.

\bibitem{Perez-Martinez:2021zjj}
R.~P{\'e}rez-Mart{\'i}nez, S.~Ramos-S{\'a}nchez, and P.~K.~S. Vaudrevange,
  ``{Landscape of promising nonsupersymmetric string models},''
  \href{http://dx.doi.org/10.1103/PhysRevD.104.046026}{{\em Phys. Rev. D}
  {\bfseries 104} no.~4, (2021) 046026},
  \href{http://arxiv.org/abs/2105.03460}{{\ttfamily arXiv:2105.03460
  [hep-th]}}.

\bibitem{GrootNibbelink:2015dvi}
S.~Groot~Nibbelink, O.~Loukas, F.~Ruehle, and P.~K.~S. Vaudrevange, ``{Infinite
  number of MSSMs from heterotic line bundles?},''
  \href{http://dx.doi.org/10.1103/PhysRevD.92.046002}{{\em Phys. Rev. D}
  {\bfseries 92} no.~4, (2015) 046002},
  \href{http://arxiv.org/abs/1506.00879}{{\ttfamily arXiv:1506.00879
  [hep-th]}}.

\bibitem{Ruehle:2020jrk}
F.~Ruehle, ``{Data science applications to string theory},''
  \href{http://dx.doi.org/10.1016/j.physrep.2019.09.005}{{\em Phys. Rept.}
  {\bfseries 839} (2020) 1--117}.

\bibitem{Silverstein:1995re}
E.~Silverstein and E.~Witten, ``{Criteria for conformal invariance of (0,2)
  models},'' \href{http://dx.doi.org/10.1016/0550-3213(95)00186-V}{{\em Nucl.
  Phys. B} {\bfseries 444} (1995) 161--190},
  \href{http://arxiv.org/abs/hep-th/9503212}{{\ttfamily arXiv:hep-th/9503212}}.

\bibitem{Anderson:2022kgk}
L.~B. Anderson, J.~Gray, M.~Larfors, and M.~Magill, ``{Vanishing Yukawa
  Couplings and the Geometry of String Theory Models},''
  \href{http://arxiv.org/abs/2201.10357}{{\ttfamily arXiv:2201.10357
  [hep-th]}}.

\bibitem{GrootNibbelink:2009wzz}
S.~Groot~Nibbelink, J.~Held, F.~Ruehle, M.~Trapletti, and P.~K.~S. Vaudrevange,
  ``{Heterotic Z(6-II) MSSM Orbifolds in Blowup},''
  \href{http://dx.doi.org/10.1088/1126-6708/2009/03/005}{{\em JHEP} {\bfseries
  03} (2009) 005}, \href{http://arxiv.org/abs/0901.3059}{{\ttfamily
  arXiv:0901.3059 [hep-th]}}.

\end{thebibliography}\endgroup
